# Cosmological Theories Of The Extra Terms

A.M. Deakin
*e-mail: amdeak.deakin@btinternet.com*

L.H. Kauffman
*Dept. of Mathematics, Statistics and Computer Science, University of Illinois at Chicago, 851 South Morgan Street, Chicago, Illinois 60680 USA*
*Tel: 773 363 5115; Fax:312 996 1491; e-mail: Kauffman@uic.edu*

## Introduction

*This paper purports to have: Introduced a new formulation of Quantum Mechanics, explained the apparent disconnect between Quantum Mechanics and General Relativity, explained the observed far field expansion of the Universe (Dark Energy), supplied an argument which goes towards explaining away Dark Matter (there are modelling difficulties) and not explained, on the basis of gravitational theory, the Voyager Anomaly.*

## 0. Concepts

Constraints Theory (CT) [22, 23] is a branch of theoretical physics. It begins with Quantum Mechanics (QM) but has connections with Classical Mechanics (CM) and Cosmology. Its original purpose was to explain why certain structures appear in CM on the basis that QM is fundamental; and why these structures are often successful as a basis for predictive/ descriptive quantum calculations about the real world. But the applications of CT are even wider.

CT is based on a formulation of QM that replaces scalar observables by Hermitian operators and differentials of scalar observables by commutators. It thereby uses the Schrodinger [1] method rather than the path integral method developed by Feynman [2]. But CT does not use the structures found in a Lagrangian or a Hamiltonian formulation of CM to construct QM; a method used originally, in different ways, by both of those authors. One has sympathy with them: For how are they to inform the problem unless they impose structure? Where is the structure to come from apart from classical Lagrangians or Hamiltonians? After all Lagrangian mechanics and Hamiltonian mechanics have been very successful in predicting/ describing how the medium to large scale Universe works.

In CT the structure comes from something *inevitable*; the quantisation of an hierarchy of *differential identities*. We quantise these by methods which are roughly what Schrodinger did with his famous hydrogen model. We thereby bring in all the baggage (of coordinates and time etc.) associated with that model. In doing so we bring in half the assumptions of CM. We can be criticised for this; but we must start somewhere! In CT we then look for recognisable structure in the relations between various operators.

CT assumes Cartesian coordinates and conjugate momenta of particles in a flat, continuous space. The space, here denote, *P* may be the ordinary 3-space of Euclid or it may be the 4-space-time of Minkowski; but it is flat. *P* may contain more than one particle; and, indeed, it may contain many particles represented as a continuous fluid. The particles in *P* are structureless points with little more than coordinates, momenta and mass assigned to them. The coordinates and momenta are assumed all to be *continuous*. The coordinates of the particles are in turn assumed to be differentiable functions of a single, continuous scalar time. This time is the proper time of a single observer and an adjacent clock both at rest at the origin.

The differential identities concern the time derivatives of a continuous, differentiable function theta. Theta is assumed to be a function of the scalar coordinates which are, in turn, assumed to be functions of the continuous time experienced by the observer. There may be more than one function theta associated with a given system of particles; there is an hierarchy of identities associated with each. The candidates for theta (operator or scalar depending on the context) are taken to be the scalar functions of the coordinates that appear in the Hamiltonian (operator or scalar depending on the context); the Hamiltonian is taken to be the *complete description* of the system. The first four of the differential identities are as follows (the $q^j$ are the coordinates of the particles) [20]:

$$\dot{\theta} = \dot{q}^j \theta_{,j}; \quad \theta_{,j} \equiv \frac{\partial \theta}{\partial q^j}; \quad \dot{\theta} \equiv \frac{d\theta}{dt}$$

$$\ddot{\theta} = \ddot{q}^j \theta_{,j} + \dot{q}^j \dot{q}^k \theta_{,jk}; \quad \theta_{,jk} \equiv \frac{\partial^2 \theta}{\partial q^j \partial q^k}$$

$$\dddot{\theta} = \dddot{q}^j \theta_{,j} + 3\ddot{q}^j \dot{q}^k \theta_{,jk} + \dot{q}^j \dot{q}^k \dot{q}^l \theta_{,jkl};$$

$$\ddddot{\theta} = \ddddot{q}^i \theta_{,i} + (4\dddot{q}^i \dot{q}^j + 3\ddot{q}^i \ddot{q}^j)\theta_{,ij} + 6\ddot{q}^i \dot{q}^j \dot{q}^k \theta_{,ijk} + \dot{q}^i \dot{q}^j \dot{q}^k \dot{q}^l \theta_{,ijkl};$$
$$i, j, k, l = 1, 2, \ldots n_c \equiv n_d n_p$$

Etc.

The Einstein summation convention is in force; and, unless otherwise stated, all indices lie in the range $[1, n_c = n_p n_d]$ where $n_p$ is the number of particles and $n_d$ is the dimension of the flat space *P*.

Quantising $\dot{\theta} = \dot{q}^j \theta_{,j}; \quad \theta_{,j} \equiv \frac{\partial \theta}{\partial q^j}; \quad \dot{\theta} \equiv \frac{d\theta}{dt}$ means replacing differentials by commutators and using the product rule to obtain

$$\frac{1}{i\hbar}(\Theta H - H\Theta) = \frac{1}{2}(\Theta_{,j} \frac{1}{i\hbar}(Q^j H - HQ^j) + \frac{1}{i\hbar}(Q^j H - HQ^j)\Theta_{,j})$$

where $H$ is the Hamiltonian (Hermitian) operator and

$$h \to H; \quad \theta \to \Theta; \quad \theta_{,j} \to \Theta_{,j} \equiv \frac{\partial \Theta}{\partial Q^j}; \quad q^j \to Q^j$$

where $\to$ means 'real observable corresponding to Hermitian operator'. Thus we systematically replace all derivatives by commutators and then the costraints are demands that the commutator representations behave according to (some of) the rules of continuum calculus. See Appendix A for an account of the first constraint and its consequences.

The constraints are operator equations in *all* the coordinate operators, *all* the momentum operators, the Hamiltonian operator and the theta operator(s). The momentum operators are, by definition, conjugate to the coordinates; the Hamiltonian operator is, similarly, conjugate to the time. A theta operator is defined as a pure function of the coordinate operators; in the position representation it reduces to a scalar function of the scalar coordinates. The constraints (quantizations) corresponding to the scalar equations above are [20]

$$\left\{H^{:j},\Theta_{,j}\right\}=\lfloor H,\Theta\rfloor;$$

$$\lfloor A,B\rfloor\equiv\frac{1}{i\hbar}(AB-BA);\quad \{A,B\}\equiv\tfrac{1}{2}(AB+BA);\quad A^{:j}\equiv\lfloor Q^{j},A\rfloor;\quad A_{,j}\equiv\lfloor A,P_{j}\rfloor;$$

$$i\hbar\delta^{j}_{k}=Q^{j}P_{k}-P_{k}Q^{j}\quad Q\text{'s commute; } P\text{'s commute}$$

$$\left\{\lfloor H,H^{:j}\rfloor,\Theta_{,j}\right\}+\left\{H^{:j},H^{:k},\Theta_{,j,k}\right\}=\lfloor H,\lfloor H,\Theta\rfloor\rfloor$$

$$\lfloor A,\lfloor B,C\rfloor\rfloor\equiv\lfloor A,B,C\rfloor;\quad \{A,B,C\}\equiv\tfrac{1}{3!}(ABC+CBA+CAB+ACB+BCA+BAC)$$

$$\left\{\lfloor H,H,H^{:j}\rfloor,\Theta_{,j}\right\}+3\left\{\lfloor H,H^{:j}\rfloor,H^{:k},\Theta_{,j,k}\right\}+\left\{H^{:j},H^{:k},H^{:l},\Theta_{,j,k,l}\right\}$$
$$=\lfloor H,H,H,\Theta\rfloor$$
$$\left\{A,B,...\right\}\equiv\left\{AB...\right\}\equiv\tfrac{1}{n!}\sum_{perm}AB...\text{ there being } n \text{ operators}$$

$$\left\{\lfloor H,H,H,H^{:i}\rfloor,\Theta_{,i}\right\}+4\left\{\lfloor H,H,H^{:i}\rfloor,H^{j},\Theta_{,i,j}\right\}+3\left\{\lfloor H,H^{:i}\rfloor,\lfloor H,H^{:j}\rfloor,\Theta_{,i,j}\right\}$$
$$+6\left\{\lfloor H,H^{:i}\rfloor,H^{:j},H^{:k},\Theta_{,i,j,k}\right\}+\left\{H^{:i},H^{:j},H^{:k},H^{:l},\Theta_{i,j,k,l}\right\}$$
$$=\lfloor H,H,H,H,\Theta\rfloor$$

where, in particular, the operators (expressed in position representation)

$$\Theta(\underline{Q})\equiv\theta(\underline{q})I;\quad \Theta_{,j}=\theta_{,j}I;\quad \Theta_{,j,k}=\theta_{,jk}I;\quad \Theta_{,j,k,l}=\theta_{,jkl}I\quad\text{etc.}$$

are pure in the $\underline{Q}$; the order of the suffices is immaterial.

The following notation is used above

$$a \to A; \quad \dot{a} \equiv \frac{da}{dt} \to \dot{A} \equiv \overset{1}{A}; \quad \frac{d^n a}{dt^n} \to \overset{n}{A}; \quad n = 1, 2, ....$$

where $'\to'$ means 'a real variable is represented by the Hermitian operator'.

$$[A,B] \equiv AB - BA; \quad \lfloor A,B \rfloor \equiv \frac{1}{i\hbar}[A,B]$$

$$[A,B,C] \equiv [A,[B,C]]; \quad [A,B,C,D] \equiv [A,[B,C,D]];$$
$$\lfloor A,B,C \rfloor \equiv \lfloor A, \lfloor B,C \rfloor \rfloor; \quad \lfloor A,B,C,D \rfloor \equiv \lfloor A, \lfloor B,C,D \rfloor \rfloor$$

$$p_j \to P_j; \quad q^k \to Q^k; \quad A_{,j} \equiv \lfloor A, P_j \rfloor; \quad A^{:k} \equiv \lfloor Q^k, A \rfloor$$

$$\{A_1, A_2, ..., A_n\} \equiv \tfrac{1}{n!} \sum_{perm} A_1 A_2 ....A_n$$

where the commas on the LHS are inserted, if need be, only for clarity. The order of the arguments in $\{.\}$ is immaterial. Notice that if an element inside any of the brackets $[.], \lfloor . \rfloor, \{.\}$ is null then the bracket is null.

If, in the hierarchy of constraints, the first holds then the Hamiltonian operator can be proved to be *quadratic* in the momentum operators with pre and post coefficients that are pure, free functions of the coordinate operators; see Appendix A. The proof involves the assumption that the coordinate and the momentum operators are continuous.

The equations of motion are, in general, complicated operator equations. But, in the *classical approximation* (all operators commute), Hamilton's classical equations (with all the momenta eliminated) have the appearance of geodesic equations in a Riemannian manifold. The dimension, $n_c$, of this space is the product of dimension, $n_d$ of $P$ and the number of particles $n_p$. The coordinates of a point in this space comprise the aggregate of *all* the coordinates of the particles in *P*; so there is but one point in this space that represents the particles in *P*; it is here denoted *X*. A generalisation of this space is useful. We denote by *C* the space of all the coordinate operators of the particles in *P*, similarly aggregated, in the *position representation*. Thus *C* is an ordinary continuous space of dimension $n_c$ with one point $X$ representing the particles in $P$. The connectivity of the space *C* can be guaranteed Riemannian only if the scale is large enough (sufficiently large for CM to work). If a fluid is represented in *P* then the dimension of *C* is, strictly, infinite.

If, in the hierarchy of constraints, the second *also* holds then the so called Theta Equation (TE) can be derived; see Appendix B. The TE is an operator equation. In the position representation this reduces to a fourth order PDE with a theta (an ordinary scalar function of the coordinates) as the dependent variable and the coordinates as independent variables. As asserted above a theta operator is taken to be *any* one of the *pure coordinate operator functions* that appear in the quadratic QM Hamiltonian. The reason for this assumption is that the Hamiltonian characterises the system; and, if the first constraint holds then, these functions fully characterise the Hamiltonian and hence the system. The TE is, therefore, an archetypal *field equation* in this theory; as derived it is valid both in QM and CM.

The coordinates used in the TE, however, may not form a Riemannian space. If , nevertheless, we use these coordinates to identify points in *C*, we *define* the TE on $C$. But, because $C$ may be a classical artefact (see above), the TE, in that case, may be only valid in CM.

The coefficients of the quadratic terms in the classical Hamiltonian are taken to describe classical gravitational forces; the linear and zero order terms are taken to describe classical EM forces. There is one exception to this: when the zero order term is used as a classical Newtonian gravitational potential. When the linear and zero order terms are omitted from the Hamiltonian the TE is called the Gravitational Theta Equation (GTE). If the GTE is thought of as defined on $C$ then, for the reasons set out above, it may only be valid in macrophysical situations.

We are, in what follows, concerned only with gravitation. But talk of gravitation implies, in General Relativity (GR) at least, curvature of space-time and, necessarily, the use of curvilinear coordinates. In general $C$ *is* curved (the fundamental tensor of $C$ is comprised of *free* functions of all the coordinates defined in *P*); so $C$, providing it is Riemannian should be able to accommodate Einsteinian gravitational theory. But there is here an apparent contradiction: *P*, by definition, is *flat*; and the coordinates of a point in $C$ cannot be the aggregate $\underline{q}$ of the coordinates of all the particles in *P* unless $C$ is also *flat*. Let us suppose, for the moment, that this is so.

Now let us introduce a *curved* Riemannian space $C'$ which has the same dimensionality as $C$ and like $C$ is continuous. Suppose that, unlike *C*, the fundamental tensor of $C'$ is comprised of free functions of *curvilinear coordinates* $\underline{x}$. This tensor can be equal to the fundamental tensor of $C$ only at a point P' in $C'$ and at the corresponding point P in C. Likewise we can satisfy $\underline{x} \equiv \underline{q}$ only at those points. If we demand, in addition, that $\underline{x} \equiv \underline{q}$ and $g'^{uv}(\underline{x}) = g^{uv}(\underline{q})$ are satisfied in the *neighbourhoods* of P and P' then we have to choose the $\underline{x}$ as *Cartesian geodesics with pole* P'. Given both these circumstances the flat space $C$ can be described as *tangential* to the curved space $C'$ at the points P in $C$ and P' in $C'$.

The TE, and hence the GTE, are valid anywhere in a flat space *C*; but, in flat *C*, the GTE has no content. If, however, the GTE is expressed in terms of the $\underline{x}$ and the fundamental tensor of $C'$ then it will be valid in the neighbourhood of P' in $C'$ (providing that the $\underline{x}$ are chosen as Cartesian geodesics with pole P'). This is one of the methods of bringing the flat space of conventional QM to be consistent with the curved space of GR. The two spaces are consistent only in the neighbourhoods of the points P in $C$ and P' in $C'$ ; but QM applies to the *physically small*.

It can be proved that a Riemannian space cannot have curvature unless its dimension is greater than three. So we can ask the question: What tensor equation, defined in a space Riemannian $C'$ of dimension greater than three, reduces to the GTE when the coordinates in $C'$ are Cartesian geodesics pole P'? The answer is the Kilmister equation [7], [8]. Because tensor equations are true in any coordinate system we may use the Kilmister equation, expressed in any convenient coordinates, to examine the local consequences of the GTE (expressed in Cartesian geodesics) holding in the neighbourhood of every point in a curved Riemannian $C'$.

It should be noted that neither the classical TE nor the classical GTE are tensor equations. So, when these equations are stated as being true in the neighbourhood of a point, particular attention should be paid to the coordinates and the metric that have been assumed.

We now, for the most part, drop the primes and recognise that Riemannian $C$ can be curved providing that we use *curvilinear* coordinates $\underline{x}$ instead of the *flat* coordinates $\underline{q}$. As stated above there is a theorem which states that if $C$ is to be curved, being Riemannian, that it must have a dimension in excess of three. Note that it is hypothesised, but not proven, that, in order to derive the form of the CM Hamiltonian, we do not need to consider any of the constraints above level two.

This paper is concerned, primarily, with the Kilmister equation. This is a classical equation and therefore applies, if it applies at all, to aspects of the cosmos which can be explained by non-quantum methods. It is an ODE of fourth order; and it is satisfied by solutions of the customary classical equations of gravity which are of second order. Therefore it has extra terms in its solutions. These extra terms must be appreciable only at cosmological distances; otherwise they would not have been missed. They are thought, at first sight, to be relevant to [3], [4] and to [5]; at any rate they must produce extra physics.

The modern picture is that the Universe is appreciably *flat*; this result is based on the statistics of the deviations from uniformity of the microwave background (roughly one in 100000). The Universe, theoretically, became transparent to radiation only about 400000 years after the Big Bang. So the truth (if it is true) of the transparency dates from that epoch.

Further at most 1/6 of the matter, *sensed by gravity*, is accounted for by that which we observe with telescopes; this is also the matter which is, roughly, accounted for by *particle physics*. A total of at least 5/6 of the matter, sensed by gravity, is *dark matter* which is *not* accounted for by the present Standard Model; this is hypothesised to be a mixture of unknown particles (which are not part of the Standard Model), invisible planets and gas and dark stars (if any). The matter sensed by gravity is only 30% of the total required to produce closure; the Universe is expanding under the influence of *dark energy*. This is variously explained by Einstein's *cosmological constant* producing *vacuum energy,* by *wimps* or by *quintessence*. We stick with vacuum energy. This paper is concerned with the Kilmister equation and with explanations of [4] and [5] although not [3].

## 1. The Classical Space $C$ Is Riemannian Provided That The Space $P$ Is Riemannian

If the first constraint, in the hierarchy of constraints, holds then can be proved to require that the operator Hamiltonian is quadratic in the $P_u$ (the Einstein summation convention is in force); see Appendix A . The proof requires the assumption that the spectra of all the coordinate and momentum operators are continuous.

(1.1)

$$H \equiv \mathrm{K}\left\{G^{uv}, P_u P_v\right\} + \left\{F^j, P_j\right\} + V; \quad Q^j \equiv q^j I; \quad G^{uv}(\underline{Q}) \equiv g^{uv}(\underline{q}) I = G^{vu}(\underline{Q});$$

$$i\hbar I \delta_{jk} \equiv Q^j P_k - P_k Q^j = -i\hbar\left(q^j \frac{\partial}{\partial q^k} - \frac{\partial}{\partial q^k} q^j\right); \quad P_j P_k \equiv P_k P_j; \quad Q_j Q_k \equiv Q_k Q_j;$$

$$F^j(\underline{Q}) \equiv f^j(\underline{q}) I; \quad V(\underline{Q}) \equiv v(\underline{q}) I; \quad u, v, j = 1, 2, \ldots n_c \equiv n_p n_d;$$

$$\{A, B\} \equiv \frac{1}{2}(AB + BA); \quad \left\{A^{uv}, B_{uv}\right\} \equiv \frac{1}{2}\left(A^{uv} B_{uv} + B_{uv} A^{uv}\right)$$

Here the position representation is used; $\underline{q}$ denotes the aggregate of Cartesian coordinates of the particles in flat $P$ ; capital letters are used for operators (thus $A$ is the Hermitian operator corresponding to the real observable $a$ ); the functions $g^{uv}(\underline{q}),\ f^{j}(\underline{q}), v(\underline{q})$ are free; $\mathrm{K}$ is a constant scalar with physical dimensions (mass) $^{-1}$ in order that $H$ has the physical dimensions of energy and the $G^{uv}$ have none.

The classical approximation (all operators commute) to (1.1) is

(1.2) $h \equiv \mathrm{K} g^{uv} p_u p_v + f^j p_j + v$ Scalar See Appendix A

This we simplify because here we are only interested in Einstein gravitation

(1.3) $h \equiv \mathrm{K} g^{uv} p_u p_v; \quad f^j = 0; \quad v = 0$

So the *classical* equations of motion, defined in *flat* $C$ and $P$, are (since Hamilton's equations are valid in CM)

(1.4a) $\dot{q}^j = \dfrac{\partial h}{\partial p_j}; \quad \dot{p}_k = -\dfrac{\partial h}{\partial q^k}; \quad \dot{a} \equiv \dfrac{da}{ds}$

where $s$ is the time variable. Eliminating the $p_k$ between (1.3) and (1.4a)

(1.4b) $\dfrac{d^2 q^j}{ds^2} + \Gamma^j_{kl} \dfrac{dq^k}{ds}\dfrac{dq^l}{ds} = 0; \quad \Gamma^l_{ij} \equiv \dfrac{1}{2} g^{lk}\left(g_{ik,j} + g_{jk,i} - g_{ij,k}\right)$

the equation of a geodesic in a Riemannian space whose interval $ds$ is defined by

(1.5) $ds^2 \equiv g_{uv} dq^u dq^v$

So classical $C$ is Riemannian if $P$ is Riemannian; but both $C$ and $P$ are flat.

The flat space $C$ is tangential to a curved space $C'$ at P in $C$ and at P' in $C'$ if

(1.6) $\underline{x} \equiv \underline{q}; \quad g'_{uv}(\underline{x}) = g_{uv}(\underline{q})$ at pole $\underline{x}$

where the $\underline{x}$ are Cartesian geodesics pole P' and $g_{uv}(\underline{q}), g'_{uv}(\underline{x})$ are the fundamental tensors of $C$ and $C'$, respectively. Results (1.4b/1.5) are classical tensor equations and therefore true in any coordinate system and any pole P'. Therefore classical $C'$ is also Riemannian.

## 2. The Classical Metrics Of P And C For Weak Gravity

We assume that the spaces $P$ and $C$ are continuous and Riemannian; indeed we have shown above that $C$ is necessarily Riemannian if $P$ and $C$ are continuous and $P$ is Riemannian. In general $P$ has the dimension $n_d > 2$ and is flat. The space $C$ has the dimension $n_c \equiv n_p n_d$, where $n_p$ is the number of particles. But, for the present, we discuss simpler scenarios: In these scenarios there is but one particle $n_p = 1$; $P$ has either the dimension $n_d = 3$ and a Euclidean metric with Cartesian coordinates

(2.1) $ds^2 = ds_0^2; \quad ds_0^2 \equiv dx^2 + dy^2 + dz^2$

or the dimension $n_d = 4$ and a Minkowski metric

(2.2) $ds^2 = c^2 d\tau^2 - ds_0^2; \quad \left|ds_0^2\right| << \left|cd\tau\right|$

Here $\tau$ is coordinate time and $c$ is the speed of light. The inequality is required for Newtonian methods to be valid.

We take curvature of $C'$ to be a symptom of gravity. As stated above it can be proved that to have curvature and be Riemannian $C'$ must have dimensionality of at least $n_c = 4$. Since it has been assumed that $n_p = 1$ then $n_d > 3$. In consequence we assume that $n_c = 4$. If, in addition, there is but one time-like coordinate (per particle) then the other three must be space-like.

We suppose that, if $P$ has the Minkowski metric then, given a single particle, the metric of $C'$ is the weak gravity perturbation of Minkowski ( $n_p = 1, n_c = 4$)

(2.3) $ds^2 = (1+2U)c^2 d\tau^2 + (-1+2U)(ds_0)^2; \quad |U| << 1; \quad |ds_0| << |cd\tau|$

where $x, y, z, \tau$ are *quasi-Cartesian* coordinates and time, and $U$ is a dimensionless function of the spatial coordinates $x, y, z$ only

(2.4) $U \equiv U(x, y, z)$

It is taken to be an invariant [9].

The metric (2.3) has a small curvature (determined by the second derivatives of $U$ ); so (2.3) is sufficient to describe a weak gravitational field from the point of view of GR [9]. We assume that $U$ increases without limit as a particle in $P$ is approached; that is (2.3) is valid except in a closed neighborhood that surrounds the particle. We restrict $U$ so that it does not depend upon time because this ensures that the force, defined by

(2.5a) $\mathbf{F} \propto -\nabla U; \quad \nabla \equiv \mathbf{i}\frac{\partial}{\partial x} + \mathbf{j}\frac{\partial}{\partial y} + \mathbf{k}\frac{\partial}{\partial z}; \quad \nabla^2 \equiv \frac{\partial^2}{\partial x^2} + \frac{\partial^2}{\partial y^2} + \frac{\partial^2}{\partial z^2}$

(in Cartesian vector notation) is *conservative*

(2.5b) $\nabla \times \mathbf{F} = \mathbf{0}$ follows from (2.5a)

It then turns out that $U$ is approximately proportional to the Newtonian potential [9,10].

In conformity with notation used elsewhere

(2.6) $x^1 \equiv x; \quad x^2 \equiv y; \quad x^3 \equiv z; \quad x^4 \equiv c\tau$ ; in $C$

So the metric (2.3) becomes

(2.7)
$$ds^2 = (1+2U)(dx^4)^2 + (-1+2U)ds_0^2;$$
$$(ds_0)^2 \equiv \sum_{J=1}^{3}\left(dx^J\right)^2; \quad |U| << 1; \quad |ds_0| << |dx^4|; \text{ in } C$$

This metric is the link between Newtonian mechanics and GR. As remarked above this link applies providing that the gravity is weak ( $|U| << 1$) and the speed of matter is small compared with $c$ ( $|ds_0| << |dx^4|$ ). We emphasise that, in this discussion, the only forces on test particles are gravitational. Result (2.7) must be regarded, from its derivation, as *classical*. NB In the metric (2.7) $U$ is approximately proportional to the Newtonian potential and has the reverse sign to [9]. In the notation of [9]

(2.8) $\Omega \cong -c^2 U$ See [9] p. 101 et seq.

## 3. Motion Of A Test Particle In Weak Gravity- The Relation Between $U$ And The Newtonian Potential

Suppose that an infinitesimal test particle is acted on by a scalar gravitational field potential $v(\underline{q}')$. Then the Newtonian Hamiltonian operator for the particle is, in the position representation,

(3.1)

$$H = \sum_{J=1}^{3}\frac{P_J'^2}{2\delta m} + V(\underline{Q}'); \quad \underline{Q}' = \underline{q}'I; \quad V(\underline{Q}') = v(\underline{q}')I;$$
$$i\hbar I\delta_{JK} = Q'^J P_K' - P_K' Q'^J; \quad P_J'P_K' = P_K'P_J'; \quad Q'^J Q'^K = Q'^K Q'^J; \quad J,K = 1,2,3$$

where $\delta m$ is the *inertial mass* of the test particle and $\underline{(.)}$ denotes aggregate (of coordinates etc.). But the coordinates and momenta (Cartesian and flat) are not necessarily the same as the coordinates and momenta $\underline{Q},\underline{P}$ referred to above; hence the primes.

Comparing (3.1) with the general case (1.1) (allowed by satisfaction of the first constraint), we see that

(3.2) $G_K^J(\underline{q}') = \frac{\delta_K^J I}{2\delta m\mathrm{K}}; \quad f^J(\underline{q}') = 0; \quad J,K = 1,2,3;$ position representation

Thus the space in which $H$ (see (3.1)) is defined has three dimensions and is truly Euclidean. Further, the spaces $P$ and $C$ are identical if $C$ is flat and $n_p = 1$. The classical Newtonian expression for the acceleration vector is

(3.3) $$\frac{\dot{p}_J}{\delta m} = -\frac{1}{\delta m}\frac{\partial h}{\partial q'^J} = -\frac{1}{\delta m}\frac{\partial v(\underline{q})}{\partial q'^J}$$

So infinitesimal test particles will be subject to this acceleration.

By contrast, in GR, Einstein asserts that an infinitesimal test particle moves on a geodesic in a Riemannian space [6]

(3.4)
$$\frac{d^2x^j}{ds^2} + \Gamma_{kl}^j\frac{dx^k}{ds}\frac{dx^l}{ds} = 0; \quad \Gamma_{ij}^l \equiv \frac{1}{2}g^{lk}\left(g_{ik,j} + g_{jk,i} - g_{ij,k}\right); \quad (.)_{,k} \equiv \frac{\partial(.)}{\partial x^k}$$

Now, referring to (2.7), a link with Newtonian theory is the condition

(3.5) $$ds \approx \left|dx^4\right|; \quad \left|dx^4\right| >> \left|ds_0\right|; \quad cdt \approx ds_0$$

where $t$ is the Newtonian time variable. With this (3.4) becomes

(3.6a) $\Gamma_{44}^J = U,_{x^J}$ See (2.7)

(3.6b) $$\frac{d^2x^J}{dt^2} \approx -c^2\Gamma_{44}^J \approx -c^2U,_{x^J}; \quad J = 1,2,3; \quad \left|U\right| << 1$$

giving, approximately, Cartesian components of classical acceleration in any weak gravitational field. But this is an expression for the components of acceleration referred to coordinates that differ from those used at (3.3). This is made obvious by comparison of the metrics (2.7) and (3.2); the latter is constant and *exactly* 3-Euclidean; and the former is variable, slightly curved and *approximately* 4-Minkowskian. The latter may be made more like the former by assuming that $P$ is flat but four dimensional

$$(3.7)\qquad \begin{aligned} ds^2 &= \varepsilon\left[-(dq'^4)^2 + \frac{1}{2\delta m\mathrm{K}} ds_0'^2\right]; \\ (ds_0')^2 &\equiv \sum_{J=1}^{3}\left(dq'^J\right)^2; \quad |U| << 1; \quad \left|ds_0'^2\right| << \left|dq'^4\right| \end{aligned}$$

with $\varepsilon \equiv \pm 1$ as the indicator. We can then transform (2.7) into (3.7) by only changing coordinates and, if necessary, $\varepsilon$.

How does $v$, the classical Newtonian potential, compare with $U$ ?; see (2.7). In comparing these two variables we are contrasting a truly Newtonian case with an approximating weak field case in GR. We can make the comparison by comparing the two acceleration vectors (3.3) and (3.6b) referred to the same coordinate system.

$$(3.8)\qquad -\frac{1}{\delta m}\frac{\partial v}{\partial q'^J} \approx -c^2\frac{\partial U}{\partial q'^J} \Rightarrow v \approx \delta m c^2 U; \quad |U| << 1 \quad \text{See (3.3/3.6b)}$$

The last step relies on the potential being unspecified within a constant.

## 4. Einstein's Equations- $n_c = 4; \quad n_p = 1$

GR is essentially a geometrical theory based on Riemannian geometry. It brings in CM by noting that a particular geometrical tensor in GR has the same zero tensor divergence, as the energy-momentum ten-sor does in CM, in the expressing the laws of mechanics. The two tensors must be proportional in order to agree with the Newton/ Poisson theory for week gravity. GR treats measurements of time on the same footing as measurements of space. This is clearly wrong, in some sense, because we can place ourselves anywhere, in space, by an act of will; but we cannot do the same for time. Measurements of time necessarily increase; and it sweeps us along with it. Macroscopically it is something to do with the relentless increase in entropy treated by classical thermodynamics; microscopically it is something to do with QM. Yet the Einstein theory of GR has withstood all the experimental and observational tests for more than 100 years.

In GR the Einstein law of gravity for empty space (i.e., between particles) is the tensor equation

(4.1a) $R_{ab} = 0; \quad a,b = 1,2,3,4; \quad R_{ab}$ is the Ricci tensor [6]

Alternatively the law can be expressed as

(4.1b) $G^u_v = 0; \quad G^u_v \equiv R^u_v - \frac{1}{2} R\delta^u_v; \quad G^u_v$ is the Einstein tensor

Only when the coordinates are those of a particle does (4.1a) break down; then the RHS is a species of delta function. That is the *matter* is concentrated in the *particle*; and the curvature is *infinite* at the particle. More generally, when some of the particles are distributed evenly and are so numerous that they can be represented by a fluid, the Einstein law is given by the equation [6]

(4.2a) $G^a_b + \chi T^a_b = 0; \quad G^u_v \equiv R^u_v - \frac{1}{2} R\delta^u_v \Rightarrow G = -R \Rightarrow R^u_v = G^u_v - \frac{1}{2} G\delta^u_v$

where

(4.2b)
$$\chi \equiv 8\pi \mathrm{G}/c^4 = 2.0761\times 10^{-43} m^{-1}kg^{-1}s^2;$$
$$c = 2.99792458\times 10^8 ms^{-1};$$
$$\mathrm{G} = 6.672(59)\times 10^{-11} Nm^2kg^{-2} \equiv 6.672(59)\times 10^{-11}\ m^3kg^{-1}s^{-2}$$

and $\mathbb{G}$ is Newton's constant with $T_b^a$ as the matter-energy-momentum-stress tensor of the fluid. Another law of gravity, that Einstein suggested but later rejected (for his purposes), is

(4.2c) $G_b^a + \chi T_b^a + \Lambda\delta_b^a = 0$ Tensor equation

where, to make (4.2c) a tensor equation, $\Lambda$ is a universal constant. Note that, given (4.2c), and given a model universe empty of ordinary matter and energy

(4.3a) $T_b^a = 0 \Rightarrow G_b^a = -\Lambda\delta_b^a \Rightarrow R_b^a = \Lambda\delta_b^a$ Einstein space

More generally, when the Riemannian space is four dimensional,

(4.3b)
$$R_v^u = G_v^u - \tfrac{1}{2}G\delta_v^u = -\chi T_v^u - \Lambda\delta_v^u + \tfrac{1}{2}(\chi T + 4\Lambda)\delta_v^u = -\chi\left(T_v^u - \tfrac{1}{2}T\right) + \Lambda\delta_v^u$$

The equation (4.2c) is subject to the identity

(4.4) $G_{b;a}^a = 0 \Rightarrow T_{b;a}^a = 0$ Requires $\Lambda$ to be constant

The last equation at (4.4) is the tensor expression for the classical mass-energy-momentum conservation laws in CM [6]. They require, in order that (4.2c) should be a tensor equation, that $\Lambda$ should be constant.

This section raises the question of sign conventions. In Section 2 we have supposed that the signature of the Minkowski metric is $-1,-1,-1,+1$ making the interval $ds$ real for speeds less than $c$. Both Eddington [9] and Spain [6] observe this convention; so, in their work, Einstein's equation (4.2c) is written as above. But in more modern work the signature of the Minkowski metric is assumed to be $1,1,1,-1$ so that, for speeds less than $c$, the interval $ds$ is imaginary. The tensor $T_b^a$, although still real, then changes sign and Einstein's equation is written

(4.5) $G_b^a + \Lambda\delta_b^a = \chi T_b^a$

We assume the Eddington/ Spain convention.

# 5. The Theta Equation, The Gravitational Theta Equation And Kilmister's Equation

If the first *two* constraints hold, in the hierarchy, then we can deduce the Theta Equation valid in QM [19]

(5.1a)
$$G^{vj}(\underline{Q})\left(G^{uk}(\underline{Q})\Theta_{,jku}(\underline{Q})\right)_{,v} = 0; \quad \text{irrespective of } F^j \text{ and } V;$$
$$\Theta_{,j} \equiv \frac{i}{\hbar}\left(P_j\Theta - \Theta P_j\right) \quad \text{See Appendix B}$$

In the position representation this reduces to the PDE (irrespective of $f^j$ and $v$)

(5.1b) $$g^{vj}\left(g^{uk}\theta_{,jku}\right)_{,v} = 0; \quad j,k,u,v = 1,2,...n_c; \quad \theta_{,jku} \equiv \theta_{,j,k,u};$$

',' denotes partial differentiation; we *choose* Cartesian coordinates. In the same representation and with the same coordinates, if we choose,

(5.2) $$\theta \equiv g^{lm} \ \forall\, l,m$$

substituted we get the Gravitational Theta Equation (GTE)

(5.3) $$g^{vj}\left(g^{uk}g^{lm}{}_{,jku}\right)_{,v} = 0; \quad \text{GTE}; \quad \text{See (1.1/1.3)}$$

It is supposed that the GTE is valid in *quasi-Cartesian geodesic coordinates* $\underline{x}$ in the neighbourhood of a pole P' in $C'$ if, at the corresponding point P in $C$ coordinates $\underline{q}$, the space $C$ is tangential to the space $C'$. That is

(5.4) $$\underline{x} = \underline{q}; \quad g'_{uv}(\underline{x}) = g_{uv}(\underline{q}); \quad \text{at P' in } C' \text{ and P in } C$$

where $g'_{uv}(\underline{x})$ is the fundamental tensor of $C'$ and $g_{uv}(\underline{q})$ that of $C$.

In theory the GTE is valid in QM. But what kind of space is *C*? The argument leading to (1.4b) shows that it is Riemannian; but that argument relies on (1.3) which is a *classical* (that is macroscopic) equation. So we can prove that the GTE applies only to a classical Riemannian space and hence to CM.

The Kilmister Equation [7], [8], derived by the late Clive Kilmister (2006) from the GTE, is a *classical* tensor equation defined on a Riemannian manifold $C'$

(5.5) $K_{ab} \equiv g^{ef}(R_{ab;ef} + \frac{2}{3}R_{ae}R_{fb}) = 0;$ $a,b,e,f = 1,2,...n_c = 4;$ see (4.2a)

where ';' denotes covariant differentiation. It is otherwise known as the K equation. The K equation reduces to the GTE at the pole of Cartesian geodesics; and, because of the choice of those coordinates, approximates the GTE in the neighbourhood of the pole. Ostensibly it applies to but a single particle. When $n_c = 4, n_p = 1$ it is called the relativistic K equation (RKE) and the GTE should be called the relativistic gravitational equation (RGTE); we shall not follow this usage, however, because the meaning should be clear by the context. Note that, given (4.3a) and (5.5),

(5.6a) $R^a_b = \Lambda g^a_b \Rightarrow R_{ab;ef} = 0;$ $\frac{2}{3}g^{ef}R_{ae}R_{fb} = \frac{2}{3}\Lambda^2 g_{ab} \Rightarrow \Lambda = 0$ see (5.1)

That is, when the model universe is truly empty (of all ordinary matter and of vacuum energy), the RKE requires that $\Lambda = 0$. More generally the RKE determines the elements of the Ricci tensor $R^u_v$ and hence the fundamental tensor $g_{uv}$ and the gravitational field. It also determines, via (4.3b), the matter tensor $T^u_v$ subject to symmetries and boundary conditions. The K equation is an alternative law of gravity. The ordinary relativistic law of gravity (see (4.1a)) satisfies it; but is of fourth order, as opposed to second order, and therefore has more solutions.

# 6. Newtonian Approximations To Tensors Given Cartesian Coordinates

There follow a number of approximations of tensor quantities using Cartesian coordinates and the Newtonian scheme with the particle at the origin. Physically it is unclear whether these *can* apply only in the solar system, in our galaxy (and by extension to models of galaxies in general) or to the Universe at large.

Given the metric (2.7) it can be shown that [9]

(6.1) $$R_{ab} \approx -\Delta_{ab}\nabla^2 U; \quad |U| << 1; \quad a,b = 1,2,3,4; \quad n_p = 1; \quad n_c = 4$$

where the Kronecker delta $\Delta_{ab}$ takes its usual meaning. So that the law of (weak) gravity for empty space is, according to the Newtonian approximation,

(6.2a) $$\nabla^2 U = 0; \quad |U| << 1 \quad \text{See (4.1a)}$$

There is another law of gravity, namely, the tensor equation

(6.2b) $$R^a_b = \Lambda\delta^a_b; \quad \Lambda \neq 0; \quad \delta^a_b = 1 \text{ if } a = b, \delta^a_b = 0 \text{ otherwise see (4.3a)}$$

where $\Lambda$ is a (small) universal constant. Since, given the metric (2.7), and the fact that $U$ does not depend on $x^4$

(6.2c)
$$R^a_b = R_{rb}g^{ra}, g^{ra} \approx -1, g^{44} \approx 1 \text{ according as } r = a < 4, \; g^{ra} = 0 \text{ otherwise}$$
$$\Rightarrow \nabla^2 U \approx \Lambda; \quad |U| << 1; \quad a,b,r = 1,2,3,4; \quad \text{See (6.1)}$$

Further

(6.3a)
$$G^i_j \equiv R^i_j - \frac{1}{2}R\delta^i_j; \quad \delta^i_j \equiv g^i_j; \quad R \equiv R^a_a; \quad \text{Definitions}$$
$$G^a_b \approx -\Lambda\delta^a_b; \quad |U| << 1; \quad \text{Newtonian approximation}$$

The identity (4.4) is therefore satisfied approximately because $\Lambda$ is constant or zero. Given the metric (2.7),

$$\Gamma^{l}_{ij,s} \equiv \left(\frac{1}{2} g^{lk} \left(g_{ik,j} + g_{jk,i} - g_{ij,k}\right)\right)_{,s} \; ; \quad \text{Definitition}$$

(6.3b)

$$\approx -\Delta_{li}\Delta_{lj}U_{,l,s}; \quad |U| << 1 \quad \text{Newtonian approximation}$$

Similarly, because $R^{a}_{b}$ is small or zero,

$$K^{a}_{b} \approx g^{ef} R^{a}_{b\,;ef} \quad \text{Small element Ricci tensor approximation}$$

(6.4)

$$= g^{ef} \left(R^{a}_{b,e,f} + \Gamma^{a}_{re,f} R^{r}_{b} - \Gamma^{r}_{be,f} R^{a}_{r}\right) \quad \text{Assuming geodesic coordinates}$$

$$\approx g^{ef} R^{a}_{b,e,f} \approx -\nabla^{2}(\nabla^{2}U) \quad \text{Newtonian approximation} \quad \text{See (6.2c)}$$

The second line of (6.4) is an expansion assuming the coordinates are geodesic; in fact the third line (see (2.5a)) requires them to be Cartesian Goedesics. So, the Newtonian approximation to the relativistic K equation, is

(6.5) $$-K^{a}_{b} \approx \nabla^{2}(\nabla^{2}U) = 0; \quad |U| << 1 \quad \text{See (6.1/6.2c)}$$

There is another way of deducing (6.5): Simply substitute from (2.7), for the $g_{uv}$, into the GTE and approximate.

In all of the approximations above we have neglected second and higher order products and powers of $U$ and its derivatives.

When the particles are numerous and continuously distributed the physical conditions that attach to Newton's theory require that

(6.6) $$T^{4}_{4} \approx c^{2}\rho; \quad T^{r}_{s} \to 0 \quad \text{either } r, s \neq 4, 4; \quad |U| << 1$$

where $\rho$ is the matter/ energy density and the other elements of $T^{a}_{b}$ effectively vanish. It follows that the Einstein law of gravity (4.2a), with these conditions, reduces to

(6.7a) $\nabla^2 U \approx -c^2 \frac{\chi}{2}\rho; \quad |U| << 1; \quad \chi \equiv 8\pi \mathrm{G}/c^4$

According to (6.2c) space is suffused with a matter/ energy $\rho_{00}$ density given by

(6.7b) $-c^2 \frac{\chi}{2}\rho_{00} = \Lambda$ There is evidence that $\Lambda$ is negative

With the definitions

(6.7c) $\Omega \equiv -c^2 U$ See (2.8)

we get

(6.8) $\nabla^2 \Omega \approx 4\pi \mathrm{G}\rho; \quad |\Omega| << c^2$ Poisson's Equation as an approximation

This is the Poisson's Equation. Here $\Omega$ is the Newtonian potential (energy per unit mass) of an infinitesimal test particle. The requirement that the Einstein law should reduce to the Newton-Poisson law determines the value of the constant $\chi$; see (4.2b/6.7b). We have used (4.2a) as the Poisson expression of Newton's law, rather than (4.2c), because in Newton's theory $\Lambda = 0$.

The proofs of Poisson's equation (6.8) given in [10] and above (depending on GR) allow us to generalise the law of gravity. So, under the Newtonian scheme, (6.7a) can be regarded as fundamental. Whatever $U$ and whatever the law of gravity (providing that gravity is a central conservative force)

(6.9) $\rho \equiv -\frac{2\nabla^2 U}{c^2 \chi}; \quad |U| << 1$

might be regarded as definition of mass density. This is not a satisfactory definition however. If $W$ is a solution of

(6.10a) $\nabla^2 W = 0$

then it follows from (6.7a) that

(6.10b) $c^2 \frac{\chi}{2} \rho = -\nabla^2 (U + W)$

In other words, according to (6.7a), the Newtonian potential does not uniquely determine the density; *but with any other law for which* $\nabla^2 W \neq 0$ it does. Moreover note that (6.5/7a) requires

(6.11) $\nabla^2(\nabla^2 U) = 0 \Rightarrow \nabla^2 \rho = 0; \quad |U| << 1 \Rightarrow |\Omega| << c^2$

By this result (6.5) requires to be a solution of Laplace's equation; this is debatable.

As we have seen in $C$, whether $C$ is flat or slightly curved, Newtonian theory is a valid approximation *almost everywhere*. The exception is as follows: In *P* the neighbourhood of a particle corresponding to a small region in $C$; we denote the aggregate of such regions by *N*; so *N* is a neighbourhood of *X*. Near *X*, in *N*, the gravitational field is assumed to rise, without limit and therefore the curvature of $C$ rises without limit. Thus, in general, Newtonian theory is not valid in *N*. The result (6.5) is, in the Newtonian approximation, an alternative to Newton's law.

When the particles are numerous and the distribution of ordinary matter-energy is sufficiently uniform the particles can be replaced by a continuous fluid in *P*. The dimension of $C$ is then, strictly, *infinite*. A metric for *C* is valid in the neighbourhood of $X$ but outside *N*, as heretofore, and must be almost flat in order that we may apply the Newtonian method.

According to GR, as we have seen, weak gravity can be characterised by a single scalar potential $U$; see (2.7). Further, $U$ can be due to many particles. If *C* is truly flat ($U = 0$) and the metric of $P$ is

(6.12)
$$ds^2 = (dq^4)^2 - ds_0^2; \quad (ds_0)^2 \equiv \sum_{J=1}^{3}\left(dq^J\right)^2; \quad \left|ds_0^2\right| << \left|dq^4\right|$$

(however many particles $P$ contains) then the coordinates of *X*, used in *C*, can be chosen as an infinite repetition of those used in *P*

(6.13a)
$$\underline{q} = \{....q^1_{\alpha-1}, q^2_{\alpha-1}, q^3_{\alpha-1}, q^4_{\alpha-1}, ....q^1_{\alpha}, q^2_{\alpha}, q^3_{\alpha}, q^4_{\alpha}, q^1_{\alpha+1}, q^2_{\alpha+1}, q^3_{\alpha+1}, q^4_{\alpha+1}....\}; \quad \alpha = 1, 2, .....n_p$$

The metric of flat $C$, using these coordinates, is therefore

(6.13b)
$$ds^2 = \sum_{\alpha=1}^{n_p}\left[(dq^4_\alpha)^2 - (ds_0^\alpha)^2\right]; \quad (ds_0^\alpha)^2 \equiv \sum_{j=1}^{3}\left(dq^J_\alpha\right)^2; \quad \left|ds_0^\alpha\right| << \left|dq^4_\alpha\right|; \quad n_p \to \infty; \quad 1 \le \alpha \le n_p$$

Dividing the metric (6.13b) by $n_p$

(6.14)
$$d\overline{s}^2 = (d\overline{q}^4)^2 - (d\overline{s}_0)^2; \quad d\overline{s}^2 \equiv \frac{ds^2}{n_p}; \quad (d\overline{s}_0)^2 \equiv \frac{1}{n_p}\sum_{\alpha=1}^{n_p}(ds_0^\alpha)^2;$$
$$(d\overline{q}^4)^2 \equiv \frac{1}{n_p}\sum_{\alpha=1}^{n_p}(dq^4_\alpha)^2; \quad (d\overline{q}^J)^2 \equiv \frac{1}{n_p}\sum_{\alpha=1}^{n_p}(dq^J_\alpha)^2; \quad J = 1, 2, 3$$

The first line of (6.14) is the metric of a Minkowskian space with coordinates $\overline{q}^1, \overline{q}^2, \overline{q}^3, \overline{q}^4$; and quantities denoted with a bar over the top are RMS values of the corresponding quantities for each of the particles in *P*.

In slightly curved $C$ the metric appropriate to the $\alpha^{th}$ particle, which in this case can be treated as an infinitesimal test particle, is

$$(6.15)\quad \begin{aligned} &(ds^{\alpha})^2 = (1+2U)(dx_{\alpha}^4)^2 + (-1+2U)(ds_0^{\alpha})^2; \\ &(ds_0^{\alpha})^2 \equiv \sum_{j=1}^{3}\left(dx_{\alpha}^J\right)^2; \quad |U| << 1; \quad \left|ds_0^{\alpha}\right| << \left|dx_{\alpha}^4\right| \end{aligned}$$

where the dimensionless potential $U$ is a function of *all* the particle coordinates. There are so many particles that, other than in the neighborhood of any particle (where the Newtonian formulae are invalid), $U$ is approximately independent of the existence of any one particle. So, summing (6.16) and dividing by $n_p$,

$$(6.16)\quad \begin{aligned} &d\overline{s}^2 = (1+2U)(d\overline{x}^4)^2 + (-1+2U)d\overline{s}_0^2; \\ &(d\overline{s}_0)^2 \equiv \sum_{J=1}^{3}\left(d\overline{x}^J\right)^2; \quad |U| << 1; \quad \left|d\overline{s}_0^2\right| << \left|d\overline{x}^4\right| \end{aligned}$$

This is (approximately) the metric of a slightly curved Minkowskian space; and quantities denoted with a bar over the top are RMS values of the corresponding quantities for each of the particles in $P$ with

$$(6.17)\quad \overline{x}^J \approx \overline{q}^J; \quad \overline{x}^4 \approx \overline{q}^4$$

It follows that when the system is composed of a fluid only we may use $n_c = 4$ although the dimension of $C$ is fact infinite.

# 7. SS Solution Of The Newtonian Approximation To The K Equation

Given $n_p = 1, n_c = 4$ the Newtonian approximation to the relativistic K equation is (6.5). Suppose there is a particle at the origin. Outside *N*, the geometry in C, although subject to slight curvature, must *approximate* the geometry in $P$. We suppose, for the present discussion, that the origins of $C$ and $P$ coincide in the particle; and we presume that its gravitational field, as sensed by test particles, is spherically symmetric (SS) and so, defined in *C*,

(7.1) $$U \equiv U(r); \quad r^2 \equiv \sum_{j=1}^{3} (x^j)^2; \quad n_c = 4, n_p = 1; \quad |U| << 1$$

The Laplacian then reduces to

(7.2) $$\nabla^2 \equiv \frac{d^2}{dr^2} + \frac{2}{r}\frac{d}{dr}$$

and (6.5) becomes

(7.3) $$\left(\frac{d^2}{dr^2} + \frac{2}{r}\frac{d}{dr}\right)^2 U = 0; \quad |U| << 1$$

A general SS solution of this ODE is

(7.4) $$U = \frac{k_1}{r} + k_2 r^2 + k_3 r + k_4; \quad |U| << 1$$ Maple 16

where $k_1, k_2, k_3, k_4$ are constants and $r$ is the distance of a test particle from the origin of *C*. In order that the condition $|U| << 1$, attached to (6.5), can be satisfied we set

(7.5) $$k_4 = 0 \Rightarrow U = \frac{k_1}{r} + k_2 r^2 + k_3 r; \quad |U| << 1$$

We take it that (7.5) refers to the dimensionless potential of a test particle distant $r$ from the origin. Because (6.5) is a Newtonian approximation (to the K equation) it is subject to the same strictures as appear at the beginning of Section 6.

The solution (7.5) is seen to include the usual Newtonian 'inverse square' contribution to the potential; but it includes also *extra terms*. To have escaped experiment these extra terms must be either very small or zero (in the solar system). We assume, in what follows, that the extra terms are *non-zero*; but that they are *small* except at huge (cosmological) distances. The extra terms may provide an approximate explanation of three recently observed and mysterious phenomena [3],[4] and [5].

One of these [4] cites very distant objects (type 1a supernova) that should, according to conventional ideas be slowing down, *as speeding up*; the radial acceleration is proportional to the distance. When Hubble found that all the galaxies where moving away from each other Einstein set his $\Lambda = 0$ because it was not needed; see (4.2c). But [4] makes use of a non-zero $\Lambda$ to describe the extra acceleration. Now the Newtonian equation correspond to the Einsteinian equation (4.2a) for weak gravity; so the extra terms $k_2 r^2 + k_3 r$ must correspond to a non-zero $\Lambda$ ; see the strictures, however, at the beginning of Section 6 .

If we define $\Lambda'$ by the tensor gravity equation

(7.6a) $$R^a_b = \Lambda' \delta^a_b$$

for a model universe empty of ordinary matter where $\Lambda'$ is a universal constant (see (4.3a)). This is a different law of gravity either from (4.1a) or the K equation (5.5). We get as the 'Newtonian' approximation, in Cartesians,

(7.6b) $$\nabla^2 U \approx \Lambda'; \quad |U| << 1$$ See (2.7/6.1/6.2c/6.3)

with an SS solution

(7.6c) $$U = \frac{k_1'}{r} + k_2' + \frac{\Lambda' r^2}{4}; \quad |U| << 1$$

where $k_1', k_2'$ are constants of integration. In order to satisfy the condition attached to (7.6c) we put

(7.7a) $$k_2' \to 0$$

The terms in $k_1'$ and $r^2\Lambda'$ must be small enough to satisfy the condition on $U$ . Result (7.6c) is the ‘Newtonian’ dimensionless potential of a SS particle at the origin. If the system is of ordinary mass $m'$

(7.7b) $$k_1' \equiv -\frac{\mathsf{G}m'}{c^2}$$

So, comparing (7.6c) with (7.5) for *small* $r$ ,

(7.8a) $$k_1 \approx k_1' \quad \text{extra terms neglected}$$

Comparing (7.6c) with (7.5) for *large* $r$

(7.8b) $$k_2 \approx \frac{\Lambda'}{4}; \quad \text{terms in } k_3, k_2', k_1' \text{ neglected}$$

We have put the word ‘Newtonian’ in inverted commas ‘’ because Newton’s theory does not include the term $\Lambda'$ in its gravitational law. According to (7.5) the radial acceleration of a test particle, for large $r$ , is

(7.9) $$-c^2U_{,r} = -c^2\left(-\frac{k_1}{r^2} + 2k_2 r + k_3\right)\Bigg|_{r\to\infty} \approx -2c^2k_2 r; \quad |U| << 1$$

See (3.6b)

Since very distant objects have a positive acceleration, which is proportional to distance, (7.9) is positive [4]. This implies that $k_2$ is physically small and negative and hence $\Lambda'$ is the same; see (7.8b). Note that the extra terms, in (7.5), have nothing to do with the slow drift of the perihelion of the planet Mercury; their dependence on range is wrong!

Result (7.5) applies to a single particle. If we accept (6.9), and regard Poisson’s equation as fundamental, the mass/ energy density of the single particle may be defined as

(7.10a)
$$\rho(r) = m\delta(r) + \frac{2}{c^2\chi}\nabla^2 U$$
$$= m\delta(r) + \frac{2}{c^2\chi}\left(\frac{d^2}{dr^2} + \frac{2}{r}\frac{d}{dr}\right)U = m\delta(r) + \frac{2}{c^2\chi}\left(6k_2 + \frac{2k_3}{r}\right)$$

where $\delta(r)$ is the SS Dirac delta which has the properties

(7.10b) $\int_0^\infty 4\pi z^2 \delta(z) dz = 1; \quad \delta(z) = 0$ except at $z = 0$ when it is infinite

Here $m\delta(r)$ is the particle density and equation (7.10a) is true only at

(7.11) $r = 0$

elsewhere it is

(7.12)
$$\frac{2}{c^2\chi}\left(6k_2 + \frac{2k_3}{r}\right)$$

This term may well be negative.

If we accept (6.9) for the particle we must accept (see (6.11))

(7.13)
$$\nabla^2\rho = \left(\frac{d^2}{dr^2} + \frac{2}{r}\frac{d}{dr}\right)\rho(r) = 0 \Rightarrow \rho(r) = \rho_1 + \frac{\rho_2}{r}$$

where $\rho_1$ and $\rho_2$ are constants of integration having the physical dimensions of density and density$\times$length respectively. So, in order that this shall be consistent with (7.5),

(7.14)
$$\rho_1 = \frac{12k_2}{c^2\chi}; \quad \rho_2 = \frac{4k_3}{c^2\chi}; \quad r \neq 0$$

Because $|U| << 1$, the solution (7.5) is not accurate, at a great distance from the origin, unless the space is almost flat. It follows that $U$ and $\nabla^2 U$ must be, in some sense, 'small' in the region of validity.

# 8. Hypothesis- $k_2$ and $k_3$ Are Proportional To The Mass Of A Small Compact Portion Of Matter

Result (7.5) applies to a particle; but, because matter is made up of particles, we may sum over the particles in a *small compact* piece of matter, in the manner of Newton, providing that the fields linearly superpose and the radii are appreciably the same

(8.1)
$$U = \frac{k_1}{r} + k_2 r^2 + k_3 r \rightarrow U = \Sigma\left(\frac{\delta k_1}{r} + \delta k_2 r^2 + \delta k_3 r\right); \quad |U| << 1; \quad r \neq 0$$

where the $\Sigma$ is over the particles of the small piece of matter. This implies that

(8.2a) $\quad k_1 \propto m, k_2 \propto m, k_3 \propto m$

where $m$ is the mass of the small piece of matter. If we identify the first term in (8.1) as the ordinary Newtonian potential,

(8.2b) $\quad k_1 \propto m$ in fact $k_1 = -\dfrac{m\mathrm{G}}{c^2}$

where $m$ is the mass of the small, compact piece of matter. So our hypothesis means

(8.2c) $\quad k_1 = \kappa_1 m; \quad k_2 = \kappa_2 m; \quad k_3 = \kappa_3 m; \quad \kappa_1 = -\dfrac{\mathrm{G}}{c^2}; \quad$ See (4.2b)

We conclude that for any small, compact piece of matter of mass $dm$ the dimensionless potential $dU$ is SS and, at radius $r$, is

(8.3) $$dU = dm\left(\frac{\kappa_1}{r} + \kappa_2 r^2 + \kappa_3 r\right)$$

We do not know the values of $\kappa_2$ and $\kappa_3$ but, evidently, they are independent of $dm$; we will assume that

(8.4) $\pm(\kappa_1 < 0, \kappa_3 > 0)$; $|\kappa_2 r^2| << |\kappa_3 r|$; $\kappa_2, \kappa_3$ probably universal.

## 9. The Theory Behind (8.3) Applied To A Flat Galaxy [5]

It is customary to apply Newtonian theory to the detailed structure of galaxies save for the absolute centre where, for some or all galaxies, there is a massive black hole and where Einsteinian theory is appropriate. This is despite the facts that galaxies are of the order of 100000 *ly* across and, in Newton's theory as opposed to Einstein's, time is universal. Probably Newton's theory works, when applied to galaxies, because the light transit time is so small compared to the age of the Universe.

By a 'flat galaxy' we mean a galaxy like our own. Judging by the observed luminosity the form includes spirals. It is a more or less thin disc with a bright ball in the centre tailing off, in brightness, thickness and density, towards the periphery. It is postulated that it has an invisible halo, centred on the galaxy, of much greater radius than the luminous part; this halo taken to approximates an oblate spheroid and can be assumed to be rotating. According to modern theories 83% to 99% of the matter in the galaxy is in the halo. This is the so called Dark Matter [5]. Dark matter may consist of dust and gas, low luminosity stars, planets (free or otherwise), neutrinos and/ or other sub-atomic particles, in particular, those that are not part of the Standard Model; or we may have simply got the gravity law wrong or both. We shall examine the hypothesis about the gravity law, in detail, for the Newtonian scheme.

The model galaxy is a simplification of our galaxy. The model galaxy rotates (not necessarily as a solid body) about an axis through the centre of the galaxy and perpendicular to its plane. It could be approximated as gas of varying (radial) density. The individual stars could be 'atoms' of that gas; and there is also 'real' gas. In practice we concentrate on the *halo*. As been said at least 83% of the matter in the galaxy is theoretically in the halo; this is a cogent simplification.

Newton proved two theorems which mean that a spherical distribution of matter, for which the density only varies with radius, may treated as a point at the centre of the sphere [10]. The law of gravity can be various as long as it is central. When the test particle is *inside* the sphere the attraction is due to the matter at a *radius less than or equal to the test particle*; when the particle is *outside* the sphere the attraction is due to a point concentrated at the centre of the same mass as the sphere of matter. We approximate the halo as a perfect sphere for simplicity. This means that, although we make an error thereby, we can ignore the luminous part of the galaxy as long as we assign the mass of the whole galaxy to the halo; see Fig. 2.

Talking of the luminous part of the galaxy: Newton's gravitational law requires that two particles, the one much heavier than the other, move in an ellipse whose focus is the position of the heavier particle [10]. The simplest form of this orbit is a circle; and yet the simulations of Fig. 3 (that use the canonical Newton's law) have at least some of the stars travelling in spirals. This fact means that the attraction is not sufficient to sustain closed orbits. The culprit is the density; it falls off with distance from the centre. This means that the Newtonian attraction falls off faster than that which is required for a closed orbit; *Bertrand's Theorem* [10] requires that the only attractions that produce closed orbits are those that vary as $r$ or as $r^{-2}$. Given Newton's law the force is proportional to $r$ when the test particle is *inside* the spherical distribution of matter and proportional to $r^{-2}$ when the particle is *outside* the distribution.

We can, in theory, apply the argument leading to (8.3) to deduce the archetype velocity curve (speed of a test particle $v$ as a function of its radius $r$). The density of the halo as a function of radius is, theoretically, required for this purpose. The article [10] calculates the connection between the parameters of a *rotating* ellipsoid of incompressible fluid. The article [19] calculates the connection between the parameters of a *stationary* spherical cloud of compressible gas. Both these calculations are complex. The halo is both rotating and a compressible gas; therefore to calculate its density is even more complex.

Our final argument, which depends on the two theorems proved by Newton, is much simpler; *but is only suggestive*. Keep in mind that, according to Newton, matter is transparent to gravitation. The result (8.3) refers to compact piece of matter of mass $dm$. If we make assumptions (8.4/8.2c), about $\kappa_1$, $\kappa_2$ and $\kappa_3$, we may integrate, approximately, over all the matter of the halo to produce (see (8.3/4)) the dimensionless potential at a point in the mid-plane of the galaxy distant $r$ from the centre. There are two regimes for the dimensionless potential: one $U1(r)$ for which $r \geq a_h$ and one $U2(r)$ for which $r \leq a_h$. Here $a_h$ is the radius of the periphery of the halo (at which the ordinary density becomes zero). We assume that the gravity is given by the Newtonian approximation to the K equation and that, in consequence, result (8.3) is satisfied:

$$U1(r) = \int_{\varepsilon}^{a_h}\int_0^{\pi}\int_0^{2\pi} g(z)\rho_h(x)x^2 \sin(\theta)d\varphi d\theta dx + \text{a constant(1)};$$

(9.1a)

$$= 2\pi\int_{\varepsilon}^{a_h}\int_0^{\pi} g(z)\rho_h(x)x^2 \sin(\theta)d\theta dx + \text{a constant(1)};$$

$$0 < \varepsilon << a_h \leq r; \quad |U1| << 1$$

$$U2(r) = \int_{\varepsilon}^{r}\int_0^{\pi}\int_0^{2\pi} g(z)\rho_h(x)x^2 \sin(\theta)d\varphi d\theta dx + \text{a constant(2)}$$

(9.1b)

$$= 2\pi\int_{\varepsilon}^{r}\int_0^{\pi} g(z)\rho_h(x)x^2 \sin(\theta)d\theta dx + \text{a constant(2)}$$

$$0 < \varepsilon << r \leq a_h; \quad |U2| << 1$$

(9.1c)

$$g(z) = \frac{\kappa_1}{z} + \kappa_2 z^2 + \kappa_3 z; \quad z^2 = r^2 + x^2 - 2rx\cos(\theta); \quad z > 0; \quad \kappa_1 = -\frac{G}{c^2},$$

where $\rho_h(x)$ is the ordinary density of the halo at radius $x$. The ordinary mass of the halo is

(9.2) $$M_h = 2\pi\int_{\varepsilon}^{a_h}\int_{0}^{\pi}\rho_h(x)x^2\sin(\theta)d\theta dx = 4\pi\int_{\varepsilon}^{a_h}\rho_h(x)x^2 dx$$

Now we approximate

(9.3a) $$\rho_h(x) = \rho_h \quad \text{a constant}$$

(9.3b) $$M_h = 4\pi\rho_h\left[\frac{x^3}{3}\right]_{\varepsilon}^{a_h} \approx \frac{4\pi a_h^3}{3}\rho_h$$

As a consequence we get (Maple 16)

(9.4a)

$$U1(r) \approx 2\pi\rho_h\int_{\varepsilon}^{a_h}\int_{0}^{\pi}g(z)x^2\sin(\theta)d\theta dx + \text{a constant}(1); \quad 0 < \varepsilon << a_h \le r$$

$$\approx \frac{1}{5}\left(5\frac{k_1}{r} + k_2\left(3a_h^2 + 5r^2\right) + k_3\left(\frac{a_h^2}{r} + 5r\right)\right) + \text{a constant}(1); \quad \varepsilon = 0$$

$$k_n = \kappa_n M_h; \quad n = 1,2,3; \quad |U1| << 1$$

(9.4b)

$$U2(r) \approx 2\pi\rho_h\int_{\varepsilon}^{r}\int_{0}^{\pi}g(z)x^2\sin(\theta)d\theta dx; \quad 0 < \varepsilon << r \le a_h; \quad |U2| << 1$$

$$\approx \frac{1}{5}\left(\frac{5k_1r^2 + 8k_2r^5 + 6k_3r^4}{a_h^3}\right) + \text{a constant}(2); \quad \varepsilon = 0$$

In view of the conditions (concerning the magnitude of the dimensionless potential) attached to (9.4a/b)

(9.4c) constant(1)=constant(2)=0

The principal approximations we have made at (9.4a/b) is that we have neglected the effects of density and the gradual reduction of thickness of the visible disc with increasing distance from the centre; both are assumed constant and the thickness and density fall to zero abruptly when the edge is reached.

The dimensionless potentials $U1(r)$ and $U2(r)$ given by (9.4a/b) look very different but when

(9.5a) $r = a_h$

they are, in fact, the same (as they should be for continuity)

(9.5b) $U1(a_h) = U2(a_h) = \frac{1}{5}\left( \frac{5k_1}{a_h} + 8k_2 a_h^2 + 6k_3 a_h \right)$ See (9.4a/b)

When

(9.6a) $r >> a_h$

(9.6b) $U1(r) \approx \frac{k_1}{r} + k_2 r^2 + k_3 r$ See (8.4/7.5)

the dimensionless potential is the same as a point, at the origin, of mass $M_h$ according to the Newtonian approximation to the K equation. As the radius decreases (subject to the condition $r \geq a_h$) the value of $U1(r)$ approaches (9.5b). When

(9.7a) $r \to 0$

the dimensionless potential is

(9.7b) $U2(r) \approx k_1 \frac{r}{a_h^3}$; $\left| \frac{8k_2 r^5 + 6k_3 r^4}{5a_h^3} \right| << \left| k_1 \frac{r^2}{a_h^3} \right|$

At first $U(r)$ is negative (we adopt positive sign outside the brackets in (8.4)), with increasing $r$, and then climbs until, for the first time,

(9.7c) $$U2(r)\big|_{r=a_h} = 0 \Rightarrow \frac{1}{5}\frac{5k_1r^2 + 8k_2r^5 + 6k_3r^4}{a_h^3}\Bigg|_{r=a_h} = 0$$

; see (9.5b/8.4). On account of (9.5b) this requires that (9.5a) should be satisfied and

(9.7d) $$U1(a_h) = 0 \Rightarrow \frac{1}{5}\left(\frac{5k_1}{a_h} + 8k_2a_h^2 + 6k_3a_h\right) = 0$$

Equation (9.7d) may be used to express $a_h$ in terms of the constants $k_n;\ n = 1,2,3$. The expressions are complex and, for the purpose of illustration, we assume

(9.8a) $$k_2 = 0$$

then

(9.8b) $$a_h = \frac{1}{6}\frac{^{+}\sqrt{-30k_1k_3}}{k_3} = \frac{1}{6}\frac{^{+}\sqrt{-30\kappa_1\kappa_3}}{\kappa_3} \Rightarrow \kappa_3 = -\frac{5}{6}\frac{\kappa_1}{a_h^2}$$ See (8.2c/8.4)

which, if we accept (9.8a) and we accept the universality the $\kappa_n$ (see (8.4)), the radii of the halos of all galaxies must be the same. So (9.8a), we must presume, does not obtain in general.

The archetype $v, r$ curve rises steeply from near zero, at a small radius, then abruptly levels out to a constant velocity (which extends to an unknown radius) [11],[12]. Fig. 1 (showing uncertainty bounds of measurements) approximates this behaviour. According to Newton's law of gravity this is impossible. According to that law beyond a certain point the velocity $v(r)$ should fall off gradually with increasing $r$; see the left hand of Fig 3. But the Newtonian approximation to the K equation (see (6.5/8.4)) is capable of producing, for a small distance beyond a certain radius, no force and hence a locally constant velocity; see (7.5).

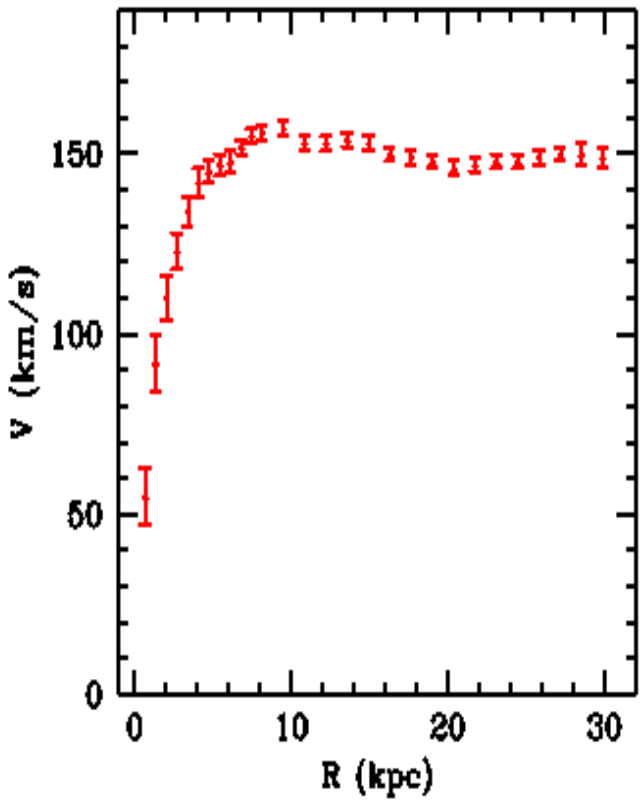

Fig.1- NGC 3198 Bergman 1989

The test particle is bound to the galaxy as long as the attractive (inwards) force is not zero. The force per unit mass of the particle $-c^2U(r),_r$ is zero at the periphery of the halo ($r = a_h$). Unfortunately neither

(9.9a) $$-c^2U1,_r\Big|_{r=a_h} = 0$$

nor

(9.9b) $$-c^2U2,_r\Big|_{r=a_h} = 0$$

is consistent with (9.7c). The equations (9.7c/d) give

(9.10a) $$k_2 = -\frac{1}{8}\frac{6a_h^2k_3 + 5k_1}{a_h^3}; \quad k_3 = k_3$$

whereas equations (9.9a/b) give

(9.10b) $$k_2 = \frac{2k_1}{a_h^3}; \quad k_3 = -\frac{15}{4}\frac{k_1}{a_h^2}$$

Result (9.10a) should give no surprise; there is effectively only one equation in two variables. Even

(9.11a) $$-c^2 U1(r),_r,_r\big|_{r=a_h} = -c^2 U2(r),_r,_r\big|_{r=a_h} = 0$$

is inconsistent with (9.10b). Equations (9.11a) give

(9.11b) $$k_2 = \frac{1}{2}\frac{k_1}{a_h^3}; \quad k_3 = -\frac{5}{4}\frac{k_1}{a_h^2}$$

The equations (9.10b) are the most important for the Newtonian theory. The others just express continuity at the edge $r = a_h$ on the plateaux of the archetype. To establish continuity, more generally, result (9.5b) should read

(9.12a) $$U1(a_h) = U2(a_h) = \frac{1}{5}\left(\frac{5k_1}{a_h} + 8k_2 a_h^2 + 6k_3 a_h\right) \neq 0$$

and result (9.11a) should read

(9.12b) $$U1(r),_r,_r\big|_{r=a_h} = U2(r),_r,_r\big|_{r=a_h} \neq 0$$

The article [12] shows that, in practice, the observed $v, r$ curve seldom, if ever, follows the archetype; not even Fig. 1 follows the archetype strictly. For example the galaxy M33, see Fig 2, has a simple observed curve; but that curve levels off to a slight rise (out to an unknown radius), when, according to the archetype, it should level off to a constant velocity at $a_h$. Others show more radical departures from the archetype; these are probably due to the effect on the dimensionless potential of the spiral arms.

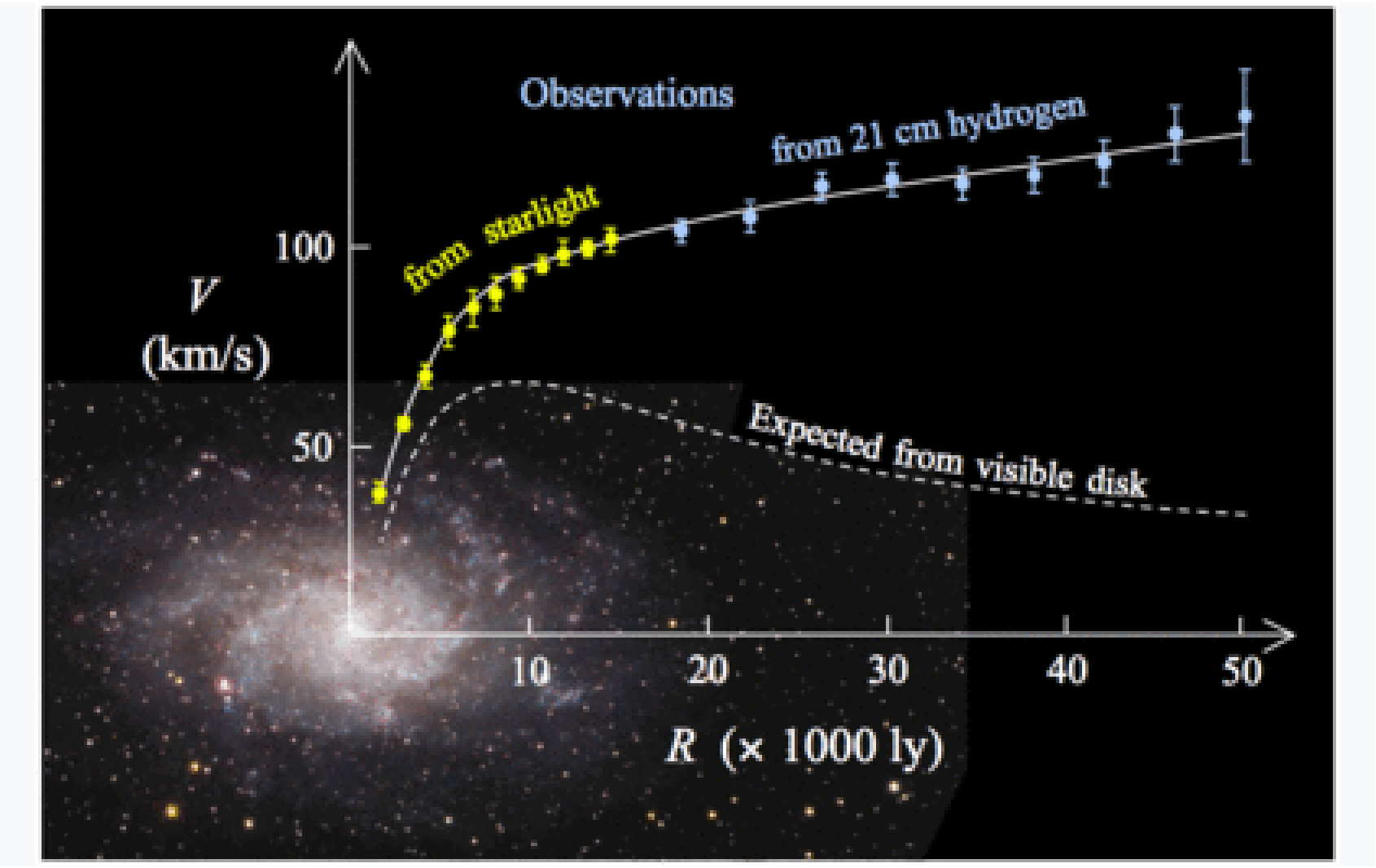

Fig 2- The $v, r$ Curve For M33

It must be concluded that the archetype $v, r$ curve is an over simplification.

It can be shown [10] that an orbit of a test particle, moving with speed $v(r)$ in a weak gravitational field characterised by an SS dimensionless potential with respect to the origin $U(r)$, satisfies

(9.13a)

$$\frac{d^2r}{dt^2} - r\left(\frac{d\vartheta}{dt}\right)^2 - c^2U,_r(r) = 0; \quad c^2 >> v^2(r) \quad \text{radial acceleration}$$

(9.13b) $$r\frac{d^2\vartheta}{dt^2} + 2\frac{dr}{dt}\frac{d\vartheta}{dt} = 0 \quad \text{transverse acceleration}$$

(9.13c) $$v(r)^2 = \left(\frac{dr}{dt}\right)^2 + \left(r\frac{d\vartheta}{dt}\right)^2; \quad v(r) > 0$$

(9.13d) $U(r) = U1(r)$ or $U2(r)$ as appropriate

where $t$ is time and $\vartheta$ is the azimuth. This gives the $v, r$ curve for our model and belongs to the Newtonian scheme although the dimensionless potential includes extra terms.

As is well known [10] elimination of the time between (9.3a/b) gives

(9.14a) $$\frac{d^2(1/r)}{d\vartheta^2} + \frac{1}{r} = -c^2 \frac{U_{,r}(r) r^2}{h_1^2}$$

where $h_1$ is a constant that satisfies

(9.14b) $$\frac{d\vartheta}{dt} = \frac{h_1}{r^2}$$

For example, when the motion is approximately circular (the presence of the $k_2, k_3$ terms in $U$ means, by *Bertrand's theorem*, that the orbit cannot be closed),

(9.14c) $$-c^2 U_{,r} + \frac{v^2}{r} \to 0 \Rightarrow v^2 \approx rc^2 U_{,r}$$

Some numerical results:

(9.15a) $k_2 = \frac{2k_1}{a_h^3}$; $k_3 = -\frac{15}{4}\frac{k_1}{a_h^2}$; $a_h = 10^5\ ly \equiv 9.4601\times 10^{20}\ m$ (assumed)

where

(9.15b)
$$\kappa_1 = -\frac{G}{c^2} = -7.4237\times 10^{-28}\ mkg^{-1};\quad k_1 = M_h \kappa_1;\quad M_h = 10^{42}\ kg$$

(9.15c)

$$k_1 = -7.4237 \times 10^{14}\ m; \quad k_2 = -1.7537 \times 10^{-48}\ m^{-2}; \quad k_3 = 3.1107 \times 10^{-27}\ m^{-1}$$

(9.15d) $$\kappa_2 = -1.7357 \times 10^{-90}\ m^{-2} kg^{-1}; \quad \kappa_3 = 3.1107 \times 10^{-69}\ m^{-1} kg^{-1}$$

We have used the halo, with the Newtonian approximation to the K equation as the law of gravity, to account for Dark Matter. If the halo does not exist the postulate of Dark Matter is still necessary in certain circumstances. To get the maximum, with Newton's law of gravity, up to the measured $v, r$ curve one has to add to the mass to the mass of the visible galaxy; even then the curve does not arrive at a true plateaux and the modelling is much more difficult. Even so, we might expect to get similar values for $\kappa_2$ and $\kappa_3$.

Another circumstance where it is necessary to take account of Dark Matter is in the space *between* the galaxies. Multiple images of quasars (predicted by Einstein's theory) show much more bending of space-time than be accounted for by the apparent (luminocity of galaxies) mean density. The conclusion must be drawn that Dark Matter permeates space.

To get an idea of how difficult is to infer the existence of Dark Matter, merely from the appearance of the galaxies, is shown by the simulations of Fig. 3.

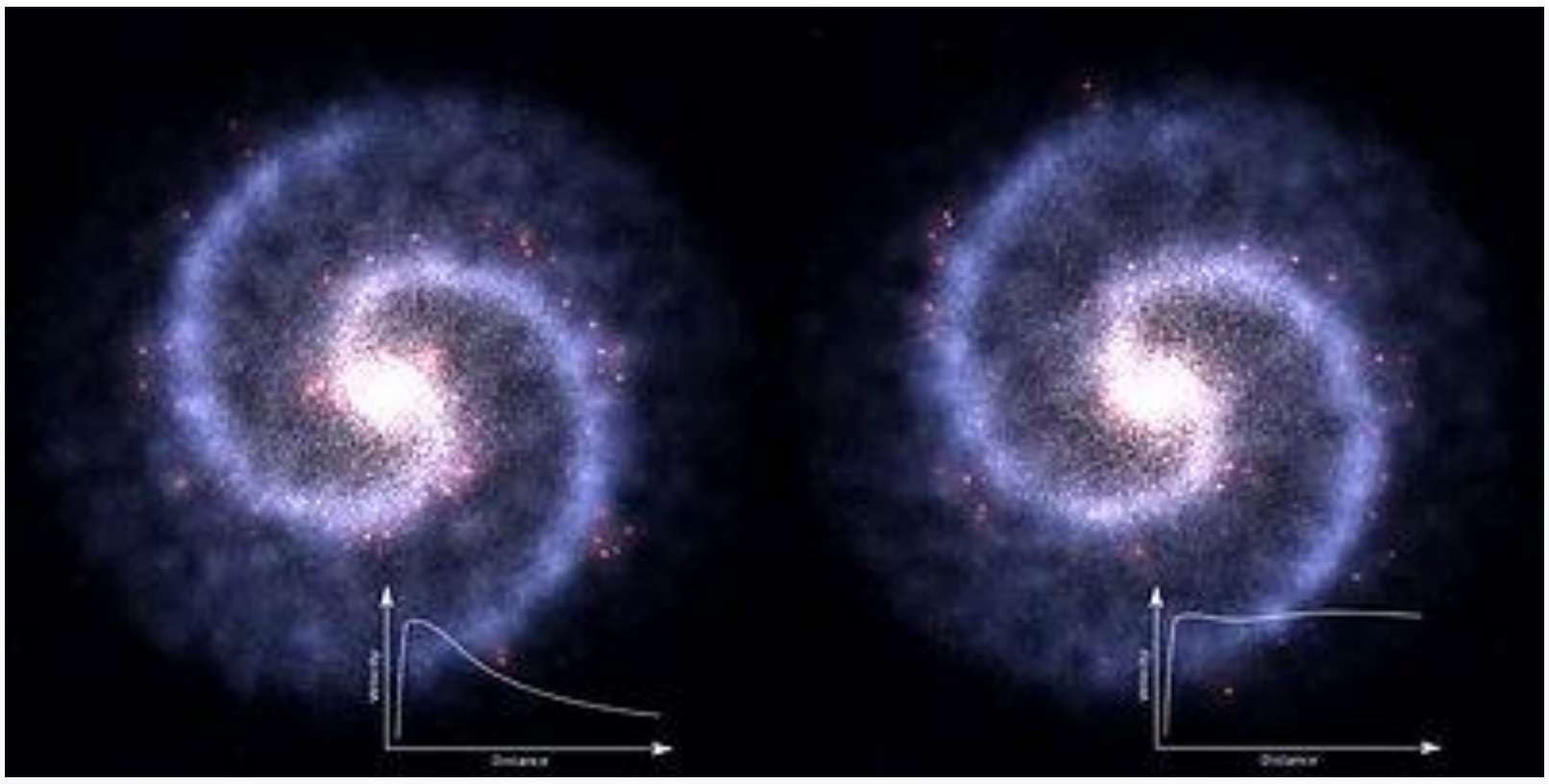

Fig. 3- Left: A simulated galaxy, with its $v,r$ curve, without Dark Matter. Right: Galaxy with an approximately flat rotation curve that would be expected under the presence of Dark Matter.

It is stated in [10] that according to the Newtonian law $G$ drifts and that it is getting smaller

(9.16) $$-\frac{1}{G}\frac{dG}{dt} \approx 3.6 \pm 1.8 \text{ parts in } 10^{11} \text{ per year}$$

wheras according to the Einstienian theory $G$ is constant. Might it be that the extra terms, in the Newtonian approximation to the K equation, roughly account for the discrepancy? To test this hypothesis we need to understand how the result is arrived at. “Observations of the occultations of fixed stars by the Moon, when corrected for all known ‘ordinary’ causes, leads to…..” result (9.16). The dimensionless potential of the Earth-Moon system is approximately

(9.17a) $$U(r) \approx \frac{k_1}{r} + k_2 r^2 + k_3 r \qquad \text{See (7.5)}$$

The force on the Moon due to the Earth is

$$(9.17b) \qquad -c^2U,_r \approx -c^2\left(-\frac{k_1}{r^2}+2k_2r+k_3\right)$$

where $r$ is the distance from the centre of the Earth to the centre of the Moon. Substituting (9.10b) into (9.17b) we get

$$(9.17c) \qquad -c^2U,_r \approx -\frac{Gm_E}{r^2}\left(1-4\left(\frac{r}{a_h}\right)^3+\frac{15}{4}\left(\frac{r}{a_h}\right)^2\right)$$

Thus the force is proportional to $G$. The Moon's radius $r$ is a function of time $t$. The dimensionless quantity $r/a_h$ is small; far too small to have any influence on (9.16). So we have the result that Newtonian law is sufficient for (9.16)

$$(9.18) \qquad -c^2U,_r \approx -\frac{Gm_E}{r^2}$$

Therefore the hypothesis is negated; the argument proceeds as in [10].

## 10. The Sun And The Pioneer Anomaly [3]

Given $r$ greater than the sun's radius we assume the form (see (7.5)) which is appropriate to a single particle at the origin

(10.1) $\quad U_o=\frac{k_{1o}}{r}+k_{2o}r^2+k_{3o}r; \quad |U_o|<<1; \quad k_{1o}=-\frac{Gm_o}{c^2}; \quad k_{2o}, k_{3o}$ constants for the gravitational potential of the sun. Here $m_o$ is the *ordinary* mass of the sun. We do not know the values to be assigned to $k_{2o}$ and $k_{3o}$ but it is reasonable to assume that for

(10.2) $r < 100\ AU$

the term associated with $k_{2o}$ can be neglected. So

(10.3) $U_o \approx \frac{k_{1o}}{r} + k_{3o}r; \quad |U_o| << 1; \quad k_{1o} = -\frac{\text{G}m_o}{c^2}$

The corresponding radial acceleration of test particles can be got from

(10.4) $-c^2 \frac{\partial U_o}{\partial r} \approx -c^2 \left( -\frac{k_{1o}}{r^2} + k_{3o} \right) = -\frac{\text{G}m_o}{r^2} - c^2 k_{3o}$

Thus, in addition to the usual inverse square term, there is a constant acceleration/ deceleration depending on $k_{3o}$. Now it is stated in [3] that at

(10.5) $r = 86\ AU$

there appeared to be a constant attraction towards the sun of

$$\begin{aligned} & c^2 k_{3o} = 10^{-10} \times \text{(acceleration due to gravity at the earth's surface)} \\ (10.6) \qquad & \approx 10^{-9}\ ms^{-2} \Rightarrow k_{3o} \approx 10^{-9} / 9 \times 10^{16}\ m^{-1} \approx 1.1 \times 10^{-26}\ m^{-1} \\ & \Rightarrow \kappa_{3o} \approx 1.1 \times 10^{-26}\ m^{-1} / 1.8 \times 10^{30}\ \text{kg} \approx 6.1 \times 10^{-57} m^{-1} kg^{-1} \end{aligned}$$

This is known as the Pioneer Anomaly. By this calculation the value of $\kappa_{3o}$ is in perfectly reasonable; but as we do not know what it *is* we are none the wiser! But, if it is to be universal, according to the calculations in Section 9 it is much too big.

There are numerous stars, in the vicinity of the sun, that, presumably, have constant components of radial acceleration/ deceleration of similar magnitude to the sun. These motions must be presumed to have all possible directions and a range of magnitudes. Will they cancel each other out? Well, if one includes the whole Universe, presumably so. But, if one includes the whole Universe, the Newtonian assumptions, on which these calculations are based, may be invalid. All we can say is that, if the Pioneer Anomaly is gravitational then, it is *not* simply related to the constant solar term.

There are many current explanations for the Pioneer Anomaly. These divide into two classes. Firstly, there are theoretical explanations which suppose that either we have the law of gravity wrong or we wrongly applying it. For example [13] considers that the time measures at the observer and at Pioneer differ, because of quantum effects, and, in consequence, the apparent motion requires correction. Secondly, there are practical explanations that suppose that we have neglected some small physical effect in evaluating the motion. For example [14] draws our attention to the non-uniform way that the structure of Pioneer radiates heat. In consequence Pioneer receives a small impulse which, it is supposed, accounts for the anomaly. The balance of opinion seems to be converging on the 'small physical effect' as the culprit.

## 11. Kilmister's Equation:-Some Analysis Concerning Isotropic-Homogeneous Space; Calculations To Do With The Cosmological Metric

We have already met Kilmister's Equation derived by the late Clive Kilmister from the GTE (see (5.3)). This is a classical tensor equation defined on a Riemannian manifold $C$

(11.1) $$K_{ab} \equiv g^{ef}(R_{ab;ef} + \tfrac{2}{3} R_{ae} R_{fb}) = 0\,; \quad a,b,e,f = 1,2,\dots n_c; \quad \text{see (4.2a)}$$

where ';' denotes covariant differentiation. It is otherwise known as the K equation. The K equation reduces to the GTE at the pole of Cartesian geodesics; and, because of the choice of those coordinates, approximates the GTE in the neighbourhood of the pole. When $n_c = 4, n_p = 1$ it is called the relativistic K equation (RKE) and the GTE should then be called the relativistic gravitational equation (RGTE); we shall not follow this usage, however, because the meaning should be clear by the context.

The RKE must be considered along with the Einstein equation (because the Einstein equation defines the mechanical tensor $T^a_b$ and so brings mass into an otherwise geometric theory); see section 4. As we have already seen (see (4.3b)), for $n_c = 4$,

(11.2)
$$R^u_v = G^u_v - \tfrac{1}{2} G\delta^u_v = -\chi T^u_v - \Lambda\delta^u_v + \tfrac{1}{2}(\chi T + 4\Lambda)\delta^u_v = -\chi\left(T^u_v - \tfrac{1}{2}T\right) + \Lambda\delta^u_v$$

When space-time is *truly empty* (of 'ordinary matter' and of 'dark' vacuum energy)

(11.3a) $$T^a_b = 0; \quad \Lambda = 0 \Rightarrow R^u_v = \Lambda\delta^u_v = 0 \Rightarrow R_{ab} = 0$$

where (11.3a) is the Einstein law of gravity in the space between particles. The RKE gives a consistent result in that it requires that the universal constant $\Lambda$ vanishes

(11.3b) $$0 + \tfrac{2}{3}(\Lambda)^2 g_{ab} = 0 \Rightarrow \Lambda = 0$$

If, however, there is no ordinary matter but there *is* vacuum energy then the law of gravity, for empty space, can still operate with $\Lambda \neq 0$

(11.4) $$T^a_b = -\frac{\Lambda}{\chi}\delta^a_b \Rightarrow R^u_v = 0 \Rightarrow G^a_b = 0,\ K^a_b = 0 \quad \text{See (4.1c/11.2)}$$

This result seems to indicate that $G^a_b$ applies to all forms of matter whereas $\Lambda$ applies to vacuum (dark) energy only.

The K equation is a collection of PDEs in the $g_{uv}$ as dependant variables and the coordinates $\underline{x}$ as independent variables. Given a fundamental tensor $g_{uv}$ which satisfies the K equation we have a Riemannian space of the points $\underline{x}$; a K space. Study of K spaces is study of the K equation. For example we may ask: Is a K space a space of constant, non-zero Riemannian curvature $\kappa$? Such a space satisfies [6]

(11.5a) $$R_{rsmn} = \kappa(g_{rm}g_{sn} - g_{rn}g_{sm}); \quad \kappa \neq 0$$

This equation implies [15] (inner product by $g^{rs}$)

(11.5b) $$R_{mn} = -(n_c - 1)\kappa g_{mn}; \quad \text{Einstein space}$$

That is a space of constant Riemannian curvature is an *Einstein space* [6] with a constant invariant. Substituting this into the K equation

(11.5c) $$0 + \tfrac{2}{3}\left((n_c - 1)\kappa\right)^2 g_{ab} = 0 \Rightarrow \kappa = 0 \Rightarrow R_{ab} = 0; \quad n_c > 1$$

So the answer to the above question is in the negative. A K space can be a space of constant Riemannian curvature but only if the curvature is zero; that is the space is flat [6],[15]. A constant curvature Riemannian space can be shown to be isotropic and homogeneous (Schur's Theorem) [15]. So the RKE does not permit the 4-space of space-time to be isotropic and homogeneous unless it is flat.

We need to consider, however, a related space. In GR the *cosmological metric* pertains to a 4-space which is *not*, in general, of constant curvature

(11.6a) $$ds^2 = d\tau^2 - \frac{A^2(\tau)}{c^2}\frac{(dx^1)^2 + (dx^2)^2 + (dx^3)^2}{\left(1 + kr^2/4\right)^2}; \quad k = -1, 0, 1$$

(11.6b) $$g_{uv} = 0, u \neq v; \quad g_{JJ} = -\frac{A^2(\tau)}{c^2\left(1 + kr^2/4\right)^2}; \quad g_{44} = 1; \quad J = 1, 2, 3$$

where units have been chosen so that the function $A(\tau)$ has the physical dimensions of length, $ds$ and $\tau$ have the physical dimensions of time and the coordinates $x^J$ have no physical dimensions [16]. The 3-subspace, formed by the $x^J$ for any given time coordinate $\tau \equiv x^4$, *is* of constant Riemannian curvature.

Going now to a physical scale which is so large, that the galaxies form individual particles of a fluid, the metric (11.6) can represent a model universe the 3-subspace of which is full of isotropic and homo-geneous matter. This is a simplification of the real Universe but it suffices for the present argument. The function $A(\tau)$ is sometimes called the *radius of the universe*. The metric (11.6) is otherwise known as the Friedman–Lemaître–Robertson–Walker (FLRW) metric [17]; it forms the basis of the Big Bang model.

We may make another interpretation of (11.6) where, for illustration, we have transformed to polars

(11.7a)

$$ds^2 = du^2 - \frac{A^2(\tau)}{\left(1+kr^2/4\right)^2}\left(dr^2 + r^2\left(d\theta^2 + \sin^2\theta d\phi^2\right)\right); \quad du^2 = c^2 d\tau^2$$

(11.7b)

$$g_{11} = -\frac{A^2(\tau)}{\left(1+kr^2/4\right)^2}; \quad g_{22} = -\frac{r^2 A^2(\tau)}{\left(1+kr^2/4\right)^2}; \quad g_{33} = -\frac{r^2\sin^2\theta A^2(\tau)}{\left(1+kr^2/4\right)^2};$$

$$g_{44} = 1$$

Here $\tau$ has the physical dimensions of time, the function $A(\tau)$ is dimensionless, $r$ has the physical dimensions of length as does $ds$ and $u$ ; and $k$ has the physical dimensions of (length)$^{-2}$. The constant $k$, in this interpretation, is continuous; it is the negative of the *Gaussian curvature* of the 3-subspace [15]; and, providing it is non-zero, scales the distance $r$. A zero value corresponds to a zero value at (11.6a); the sign for $k \neq 0$ also corresponds to the sign at (11.6a). The quantity $A(\tau)$ is often called the *expansion/ contraction factor* of the metric (11.7); that is the 3-subspace of the $r,\theta,\varphi$ expands/ contracts with coordinate time $\tau$ unless $A(\tau)$ is constant. We assume

(11.7c) $\quad A(\tau) > 0$

Given (11.7a) the tensors $R^u_v, G^u_v$ and $K^u_v$ turn out to be diagonal. Two of the four diagonal elements, in each case, are unique. So, for example,

(11.8a) $$G^J_J = -\frac{1}{A^2}\left(k + A'^2 + 2AA''\right) = -\chi T^J_J - \Lambda; \quad J = 1,2,3; \quad A \neq 0$$

(11.8b) $$G^4_4 = -\frac{3}{A^2}\left(k + A'^2\right) = -\chi T^4_4 - \Lambda; \quad G^u_v = 0, u \neq v; \quad A \neq 0$$

(11.9a) $$K^J_J = \frac{1}{3}\frac{1}{A^4}\begin{pmatrix} 4A'^2k - 4A'^4 - 7AA'^2A'' + 15A^2A'A''' \\ +11A^2A''^2 + 3A^3A'''' - 4AA''k + 8k^2 \end{pmatrix} = 0; \quad A \neq 0$$

Maple 12

(11.9b)
$$K^4_4 = \frac{3}{A^4}\left(A^2A'A''' + 4A'^2k + A^2A''^2 - 5AA'^2A'' + A^3A'''' + 4A'^4\right) = 0;$$
$$K^u_v = 0, u \neq v; \quad A \neq 0$$

where

(11.9c) $$(.)' \equiv \frac{d(.)}{du}; \quad u \equiv c\tau$$

The only check on (11.8) I have been able to make is that the expression for the Einstein tensor, given by the machine, agrees with that in [16] which was written before electronic computers existed!

The two ODEs (11.9a/b) must have consistent solutions. If they have such then it is *possible* for the K equation to be satisfied; but there is no guarantee that the metric (11.7) always pertains to a K space. That $G^u_v$ and $K^u_v$ at (11.8/9) is independent of the coordinates $r,\theta,\phi$ is a symptom of the fact that the 3-subspace is isotropic and homogeneous. In fact the transformation to polars at (11.7a), in the 3-subspace, is nugatory.

The elements of the mechanical tensor $T^u_v$ and the scalar $\Lambda$ determine the elements of the tensor $G^u_v$; see (4.2c). As has already been shown (see (11.4)), in a model universe empty of all ordinary mass/ energy but with $\Lambda \neq 0$,

$$T_b^a = -\frac{\Lambda}{\chi}\delta_b^a \Rightarrow G_b^a = 0$$

(11.10)
$$\Rightarrow G_4^4 = \frac{3}{A^2}\left(k + A'^2\right) = 0;$$
$$G_J^J = \frac{1}{A^2}\left(k + A'^2 + 2AA''\right) = 0$$

On the assumption that $k$ is constant the first of these ODEs, consistent with the second, gives

(11.11) $A(u) = \pm\sqrt{-k}u + a_0 = \frac{u}{c}a_1 + a_0; \quad a_1 = \pm\sqrt{-c^2k}$ Maple 12 and manual

where $a_1$ and $a_0$ are constants of integration. NB It is more convenient here and in the sequel to express $A$ as a function of the length $u$ rather than the time $\tau$. If $a_1$ is to be real then

(11.12a) $k \leq 0$ characteristic of an empty model universe

The Gaussian curvature of a closed 3-subspace is positive. This means that the 3-space of the empty model universe is open (hyperbolic) unless it is flat. The constant $a_0$ can be determined by the initial condition

(11.12b) $A = 1$ when $\tau = 0 \Rightarrow u = 0 \Rightarrow a_0 = 1$

This condition assumes that at the instant 'now' is the origin of time $\tau = 0 \Rightarrow u = 0$ and that the model universe is not expanded at that instant. So, finally, either

(11.13a) $A(u) = \pm\sqrt{-k}u + 1$ or $A(u) = 1 \,\forall\, u; \quad k \leq 0$

and *both* the ODEs (11.8) are exactly satisfied. Further Maple 16 shows that *both* the ODEs (11.9) are exactly satisfied by this solution. With the metric (11.7) the model universe, empty of ordinary matter, either expands or contracts, in proportion to the coordinate time at every point, or it is stationary (in, as it turns out, unstable equilibrium). Alternatively, we can shift the origin of time $\tau = 0$ to *the beginning*, providing that

(11.13b) $A(0) = a_0$; $a_0$ may be zero but if $a_0 \neq 0$ then $1 >> |a_0|$

We note a remark in [18] that 'we can have curvature without matter but not matter without curvature'.

But an empty model universe, even one with the cosmological constant non-zero, is of limited interest! To go further we begin by introducing the mean pressure $p(u)$ and the mean density of ordinary mass $\rho(u)$ (averages taken over space $x^J$, $J = 1,2,3$ but not too far!) as functions of time/ distance $u$; These quantities can be defined [16], in the case (11.7), by

(11.14) $$p_J \equiv -T_J^J; \quad c^2\rho \equiv T_4^4; \quad J = 1,2,3; \quad T_b^a = 0, a \neq b$$

where both $p_J$ and $c^2\rho$ have the physical dimensions of energy per unit volume. In general, by virtue of the Einstein equations (4.2c),

(11.15a)
$$G_J^J = -\frac{1}{A^2}\left(k + A'^2 + 2AA''\right) = -\chi T_J^j - \Lambda = \chi p_J - \Lambda; \quad J = 1,2,3; \quad A \neq 0$$

(11.15b) $$G_4^4 = -\frac{3}{A^2}\left(k + A'^2\right) = -\chi T_4^4 - \Lambda = -\chi c^2\rho - \Lambda \quad \text{See (11.14)}$$

If we consider the special case (a model universe empty of ordinary matter) we get

(11.15c)
$$\begin{array}{l} G_b^a = 0 \Rightarrow \text{pressure} \equiv -T_J^J = \dfrac{\Lambda}{\chi} \equiv p_{00J} \text{ (say)} = p_{00}; \\ c^2 \times \text{density} \equiv T_4^4 = -\dfrac{\Lambda}{\chi} \equiv c^2\rho_{00} \text{ (say)} \end{array} \quad \text{See (11.10)}$$

Note that in Section 7 we have already shown that, in the 'Newtonian' case in order to agree with observation, $\Lambda$ is negative; so $\rho_{00}$ is positive and $p_{00J} \equiv p_{00}$ is negative in a model universe empty of ordinary matter.

We now introduce Hubble's constant [16/18]

$$(11.16) \qquad \mathrm{H}_0 = \underset{u \to 0}{Lt} \frac{\dot{A}(u)}{A(u)} = c \underset{u \to 0}{Lt} \frac{A'(u)}{A(u)} \approx 2.055 \times 10^{-18}\ s^{-1}$$

That is, wherever we put the origin ($u = 0$), the model universe and the actual Universe seems to expand (see (11.13a)). From (11.15b) we deduce Friedmann's equation

$$(11.17a) \quad k + \frac{\mathrm{H}_0^2}{c^2} = \frac{\chi c^2 \rho_0 + \Lambda}{3} \qquad \text{See (7.9a)}$$

because, by definition,

$$(11.17b) \quad A(0) = 1 \qquad \text{See (11.12b)}$$

Transposing (11.17a)

$$(11.17c) \quad \mathrm{H}_0^2 = c^2 \left( \frac{\chi c^2 \rho_0 + \Lambda}{3} - k \right)$$

In other words Hubble's constant increases with the mean ordinary density $\rho_0$ and

$$(11.18) \qquad \mathrm{H}_0(\min) = +c\sqrt{\frac{\Lambda}{3} - k}; \quad \rho_0 = 0$$

Since, in an empty model universe, $k < 0$ and $\mathrm{H}_0(\min)$ is, by definition, real (see (11.11))

(11.19) $\frac{\Lambda}{3} - k > 0 \Rightarrow -3k < -\Lambda \Rightarrow 3k > \Lambda$ in an empty model universe

## 12. Here And Now Various Quantities Are Small So We Might Seek A Perturbation

We continue the argument for $\rho_0 > 0$ by perturbing the solution (11.11/13a) on the grounds (perhaps spurious) that, here and now, $c^2 \chi \rho_0 / k, \chi p / k$ are small compared to unity. Many astronomers believe, however, that $k = 0$; in which case the results are spurious. If we do not make this assumption we have some hope!

We have

(12.2a) $$A(u) \equiv 1 + \frac{u}{l} + \varepsilon f(u); \quad 0 \neq |\varepsilon| << 1; \quad k \neq 0 \equiv -\frac{1}{l^2}{}_1$$

Take the positive sign in $\frac{1}{l} = \pm\sqrt{-k}$ to give

(12.2b) $$A'(u) = \frac{1}{l} + \varepsilon f'(u); \quad A''(u) = \varepsilon f''(u);$$
$$A'''(u) = \varepsilon f'''(u); \quad A''''(u) = \varepsilon f''''(u)$$

where $\varepsilon$ and function $f(u)$ are defined as having no physical dimensions. It is to be understood that the small density and pressure, here and now, is the perturbing agency and that is small, compared with unity, but not zero. Substitute (12.2b), for , into (11.15b)

$$\frac{3}{A^2}\left(k+A'^2\right)=\frac{3}{\left(1+\varepsilon f(0)\right)^2}\left(-\frac{1}{l^2}+\left(\frac{1}{l}+\varepsilon f'(0)\right)^2\right)$$

(12.3) $$\approx 6\left(1-2\varepsilon f(0)\right)\frac{2}{l}\varepsilon f'(0)\approx 12\frac{\varepsilon f'(0)}{l}=\left(\chi c^2\rho_0+\Lambda\right)$$

$$\Rightarrow \varepsilon f'(0)\approx\frac{l}{12}\left(\chi c^2\rho_0+\Lambda\right);\quad f'(0)\neq 0$$

neglecting the terms in $\varepsilon^2$ and higher powers. Is $\varepsilon f'(0)$ small enough to be regarded as a perturbation of ? We opened this section with an assertion that was small (compared to unity); small enough to amount to perturbation of the uniform motion (11.11) characteristic of an empty universe. Well, utilising (11.17a),

(12.4)

is to be compared with Some astronomers believe

On the other hand, is free, so we can choose

(12.5a)

which, according to (12.4), makes as small as we like consistent with the approximation made at (12.4). If we make this choice, however, (11.17a) requires

(12.5b)

That is, the ordinary mean mass/ energy density is approximately constant and universal. This result is reminiscent of (11.15c) and corresponds to an empty model universe with

(12.5c)

So, since , the values given at (12.6a) for and hence are *spurious*. If we proceed with the formal perturbation, however, we get from (11.15a/b)

(12.6)

In other words an expanding model universe empty of ordinary matter! See (11.10/11/15).

## 13. Finite Expansion Of $A(u)$ As An Approximate Solution

There might be no perturbation of the solution (11.13a); if there is, all we have proved above is that, the perturbation is stable. Maple 16 can solve equations (11.8) but the solutions are immensely complex and implicit; and I cannot identify the constants of integration.

We try a Taylor's expansion about the origin 'here and now' $u = 0$

(13.1a)

$$A(u) \equiv a_0 + a_1 u + a_2 u^2 + a_3 u^3 + a_4 u^4 + a_5 u^5 + O(u^6); \quad u \geq 0$$

$$A(0) \equiv 1 \Rightarrow a_0 = 1; \quad \text{Initial Condition}; \quad 1 >> |\mathrm{u}|; \quad \text{Approximation}$$

where the coefficients $a_0, a_1, ...., a_5$ are free. The second line of (13.1a) derives from the initial condition 'here and now'. Substituted into all four of the equations (11.8a/b) and (11.9a/b) this gives four algebraic equations for the coefficients $a_1, a_2, ....a_4$; (the coefficient $a_5$ happens to vanish). This gambit solves the problem that both the equations (11.9a/b) having the same solution $A(u)$. In fact the distance/ time $u$ which we actually measure is negative. So the $A'(0)$ is negative. We might as well admit this by defining

(13.1b)

$$A(u) \cong a_0 - a_1 u + a_2 u^2 - a_3 u^3 + a_4 u^4 - a_5 u^5 + O(u^6); \quad u \leq 0$$

$$A(0) \equiv 1 \Rightarrow a_0 = 1; \quad \text{Initial Condition}; \quad 1 >> |\mathrm{u}|; \quad \text{Approximation}$$

In practice it makes no difference, to the results, which we choose of the definitions (13.1).

Because the density and pressure vary away from the origin we should write, at the least,

(13.2) $$p \cong p_0 + p_1 u + O(u^2); \quad \rho \cong \rho_0 + \rho_1 u + O(u^2); \quad |u| \to 0$$

as an approximation. This assumption will probably not be valid very far into the past or the future but it is simple. In consequence substitution of definitions (13.1/2) into equations (11.8/9) produces four polynomial equations in $u$; the coefficients of the powers of $u$ in these four equations are functions of the eleven constants $a_0, a_1, ....a_4, k, p_0, p_1, \rho_0, \rho_1, \Lambda$. If we wish to solve for these constants we must produce more equations. In explanation, we get from the initial condition,

(13.3a) $$a_0 = 1 \quad \text{See (13.1)}$$

and from the definition of Hubble's constant

(13.3b) $$a_1 = \frac{H_0}{c} \quad \text{See (11.16)}$$

so that leaves nine constants $a_2, a_3, a_4, p_0, p_1, \rho_0, \rho_1, k, \Lambda$. The object is to produce nine equations in the unknowns $a_2, a_3, a_4, p_0, p_1, \rho_0, \rho_1, k, \Lambda$. The first seven are variable, depending on where we put the origin $u = 0$; the last two, $k$ and $\Lambda$, are universal constants by definition. We can get more putative equations by any one of three alternative procedures:

*Equating coefficients, of powers $u^2$ and higher, on both sides of some of the original polynomial equations.*

*Covariantly differentiating the original tensor equations (4.2c) and (5.5) and setting up more polynomial equations via the metric (11.7a/b).*

*Going back to the QM hierarchy of constraints and postulating that the third or higher constraint holds, deriving the equivalent classical formula, eliminating the $p_k$, and converting this to a tensor equation, on the assumption that the equivalent classical formula uses Cartesian geodesics, and setting up more polynomial equations via the metric (11.7a/b).*

Clearly, these procedures are in ascending amounts of work. The work required by the last is enormous; unfortunately, from the point of view of the present theory, it is probably the only entirely valid one.

The first procedure has been tried; it is easily accomplished by Maple 16. It forces particular constraints on the differentials at $u = 0$; and these may not apply in practice. A similar criticism may be made of the second procedure; but at least it manipulates tensor equations. The first produces very high densities and pressures; this is entirely wrong for most of the history of the Universe.

Using the original five equations, including (13.3b), Maple 16 has given formulae for $a_1, a_2, a_3, a_4, k$ in terms of the rest of the quantities $p_0, p_1, \rho_0, \rho_1, \Lambda$. As is to be expected these formulae do not include the quantities $p_1$ and $\rho_1$; the five equations do not include $a_5$ either. The formulae show that there is a solution for a model universe empty of ordinary matter

$$
\begin{aligned}
& p_0 = \frac{\Lambda}{\chi}, \rho_0 = -\frac{\Lambda}{\chi c^2} \\
& a_0 = 1 \\
(13.4) \qquad & a_1 = \frac{H_0}{c} = 6.8548 \times 10^{-27}\ m^{-1} \\
& a_2 = 0, a_3 = 0, a_4 = 0 \\
& k = -\left(\frac{H_0}{c}\right)^2 = -4.6988 \times 10^{-53}\ m^{-2}
\end{aligned}
$$

As previous work has shown this solution is exact (see (11.13a)). The last line of (13.4) corresponds to (11.15c).

Many astronomers believe that $k = 0$. That being so the formulae show that we have only to supply values for $p_0$ and $\rho_0$ to solve for $\Lambda$. The assumptions are shown first; then the results

(13.5a) $\qquad k = 0; \quad p_0 = 0; \quad \rho_0 = 8 \times 10^{-27}\ kgm^{-3}$ closure value of density

(13.5b)
$$\begin{aligned} a_0 &= 1 \\ a_1 &= 6.8548\times10^{-27}\ m^{-1} \\ a_2 &= -1.3823\times10^{-53}\ m^{-2} \\ a_3 &= 8.0244\times10^{-80}\ m^{-3} \\ a_4 &= -8.0798\times10^{-106}\ m^{-4} \\ \Lambda &= -8.3052\times10^{-54}\ m^{-2} \end{aligned}$$

follow. Note that $\Lambda$ is negative as required, by the Newtonian case, to produce the observed far field positive acceleration; see (7.11) to (7.14).

The universal constant $k$ is, in the above calculation, on the cusp of becoming positive. Another calculation shows that it can become positive according to the formulae. We simply give it a positive value and repeat the calculation that leads to (13.5).

(13.6a) $$k = 10^{-54}\ m^{-2};\quad p_0 = 0;\quad \rho_0 = 8\times10^{-27}\ kgm^{-3}$$

(13.6b)
$$\begin{aligned} a_0 &= 1 \\ a_1 &= 6.8548\times10^{-27}\ m^{-1} \\ a_2 &= -1.3323\times10^{-53}\ m^{-2} \\ a_3 &= 8.0892\times10^{-80}\ m^{-3} \\ a_4 &= -8.0486\times10^{-106}\ m^{-4} \\ \Lambda &= -5.3052\times10^{-54}\ m^{-2} \end{aligned}$$

Such a positive value of $k$ may be attributed to matter and hence gravitation. The coefficients in the series for $A(u)$ are hardly changed; but $\Lambda$, while remaining negative and therefore legitimate according to the observations [4], is appreciably changed. See Friedmann's equation (11.17a).

It is to be expected that the series (13.1), for $A(u)$, will converge nicely for small values of $u$. The series are hardly convergent for

(13.7a) $u = |u|_{\max} = \tau_0 c$

where $\tau_0$ is the age of the Universe which is inferred from measurements to be

(13.7b) $\tau_0 = 13.8 \times 10^9$ years

so

(13.7c) $|u|_{\max} = \tau_0 c = 1.3054 \times 10^{26}\ m$

We can conclude that the values given for $A(|u|_{\max})$ by (13.1/6b) are very approximate. This is underlined by the fact that, in this theory, the age of the model universe is given by (keeping the same origin)

(13.8) $A(\tau_0 c) = 0$

where, in this case, (13.8) is a quartic with real coefficients; we take only the solution which is real and positive (if any). Corresponding to (13.6b) we get

(13.9) $\tau_0 = 2.6429 \times 10^{10}$ years

which is much too large. So we cannot calculate the age of the model universe by this method.

We can translate the origin to the beginning, or close to the beginning (when, according to current theories, quantum effects take over about 400000 years from the Big Bang). But we do not know the correct form to give $A(u)$ and we do not know which values to give $p_0$ and $\rho_0$. All we know is that $\min(A(u))$ is probably not zero. A more accurate specification of (13.8), referred to the 'here and now' origin, is probably

(13.10) $A((\tau_0 + 4 \times 10^5)c) = 0; \quad \tau_0$ in years

which would make hardly any difference to the final figure. But, if we are willing to accept the current figure for the age of the Universe, (13.8) gives another equation for the $a_2, a_3, a_4$. With this we can solve for another variable. We choose $\rho_0$. The calculation becomes

(13.11a) $k = 10^{-54}\ m^{-2};\quad p_0 = 0;\quad \tau_0 = 13.7\times 10^9$ years

$$a_0 = 1$$
$$a_1 = 6.8548\times 10^{-27}\ m^{-1}$$
$$a_2 = -6.3932\times 10^{-53}\ m^{-2}$$
(13.11b) $$a_3 = -9.6335\times 10^{-78}\ m^{-3}$$
$$a_4 = -2.1436\times 10^{-105}\ m^{-4}$$
$$\Lambda = -2.0774\times 10^{-52}\ m^{-2}$$
$$\rho_0 = 1.8849\times 10^{-26}\ kgm^{=3}$$

The full results, corresponding to (13.4) to (13.11), is given in Appendix C. Some small explanation is in order. The reader will see that from the results, from time to time, we have set up the polynomials for $A(u)$, $p$ and $\rho$; the polynomials for $o1$ and $o4$ correspond to (11.9a) and (11.9b) and those for the ordinary density $den$ and pressure $pres$ correspond to (11.15a) and (11.15b). We have only printed these once; although we have had to regenerate them three times 'behind the scenes'. We give the versions for $u = 0$ because they are the equations that are actually solved for various purposes.

Summarising: The first solution gives the full formulae for $a_1, a_2, a_3, a_4, k$ in terms of $\Lambda, p_0, \rho_0, H_0, c$ and $\chi$; see Appendix C. The second solution gives the formulae for the empty model universe. The third solution gives numerical values. The fourth solution brings in the extra equation (13.8/1a) and gives numerical values. For the calculation (13.11) we have chosen the second solution (enclosed in brackets {}) because that has $\Lambda$ negative and $\rho_0$ positive. The universal constant $\Lambda$ has increased in magnitude over the value given in (13.6b) and the ordinary density $\rho_0$ likewise. The ordinary density value assumed at (13.6a) is approximately the closure value.

# Appendix A- If Level One Of The Hierarchy Is Satisfied

## A1. Quantisation

The first level in the scalar hierarchy is

(A1.1) $\dot{\theta} = \theta,_j \dot{q}^j; \quad j = 1, 2, ....n_c \equiv n_d n_p;$ Einstein convention in force where $n_d$ is the dimension of the flat space where quantum phenomena take place, $n_p$ is the number of particles in that space, $\theta$ is an arbitrary pure, real function of *all* the real coordinates $q^1, q^2, ...., q^{n_c}$ and

(A1.2) $$\theta,_j \equiv \frac{\partial \theta}{\partial q^j}$$

The dot denotes differentiation with respect to time. All variables are continuous.

Quantising (A1.1), replacing differentials by commutators and using the product rule (see below (A1.6b))

(A1.3a) $$\frac{1}{i\hbar}(\Theta H - H\Theta) = \frac{1}{2}(\Theta_{,j} \frac{1}{i\hbar}(Q^j H - HQ^j) + \frac{1}{i\hbar}(Q^j H - HQ^j)\Theta_{,j})$$

where $H$ is the Hamiltonian (Hermitian) operator and

(A1.3b) $$h \to H; \quad \theta \to \Theta; \quad \theta_{,j} \to \Theta_{,j} \equiv \frac{\partial \Theta}{\partial Q^j}; \quad q^j \to Q^j$$

where $\to$ means 'real observable corresponding to Hermitian operator'. The operator $\Theta$ is pure in all the coordinate operators $Q^j$. The scalar $\hbar$ has the approximate value

(A1.3c) $\hbar = \frac{6.626075\times 10^{-34}}{2\pi}\ Js = 1.054573\times 10^{-34}\ Js$

The differentiation at (A1.3b) is purely formal and algebraic. Note that also that, with this proviso, and restricting the operators $\Theta$ and $H$ to a polynomials

(A1.4a) $\Theta_{,j} \equiv \frac{\partial\Theta}{\partial Q^j} = \frac{1}{i\hbar}(\Theta P_j - P_j\Theta);\quad H^{,j} \equiv \frac{\partial H}{\partial P_j} = \frac{1}{i\hbar}(Q^j H - HQ^j)$

where

(A1.4b) $Q^j P_k - P_k Q^j = i\hbar\delta_k^j I;\quad P_j P_k = P_k P_j;\quad Q_j Q_k = Q_k Q_j;\quad p_j \to P_j$

So, in a more compact notation (A1.3a) can be written,

(A1.5) $\frac{1}{i\hbar}(\Theta H - H\Theta) = \frac{1}{2}(\Theta_{,j}H^{,j} + H^{,j}\Theta_{,j})$

where remember that $\theta$ is arbitrary and $\theta \to \Theta$.

We need to introduce more notation:

(A1.6a)
$$AB - BA \equiv [A,B]$$
$$\frac{1}{i\hbar}[A,B] \equiv \lfloor A,B \rfloor;\quad AB + BA \equiv (A,B);$$
$$\frac{1}{2}(AB+BA) \equiv \{A,B\} = \{AB\}$$

where $A$ and $B$ are any linear operators whatsoever. Generalising (A1.6a)

(A1.6b)
$$\frac{1}{(i\hbar)^2}(A(BC-CB)-(BC-CB)A)) \equiv \frac{1}{(i\hbar)^2}[A,B,C] \equiv \lfloor A,B,C \rfloor;$$
$$\{A,B,C,D....\} \equiv \{ABCD\} \equiv \frac{1}{n!}\sum_{perm} ABCD$$

where the product rule (second line of (A1.6b)) applies to quantum mechanics

(A1.6c) $a \to A; \quad b \to B; \quad c \to C; \quad d \to D; \quad abcd \to \{A, B, C, D\}$

For example (A1.5) can be written

(A1.7) $$-\lfloor H, \Theta \rfloor = \{\Theta_{,j}, H^{,j}\}$$

The commas are redundant in brackets of type $\{\ \}$. If any of the operators are zero in brackets of the types $\lfloor\ \rfloor, \{\ \}, [\ ]$ then the bracket is zero.

The above notation as it appears in (A1.4) needs investigation: Let $X$ be an Hermitian polynomial mixture of all the coordinates $Q^j$ and all the momenta $P_k$ that is

(A1.8) $$X \equiv X_1 + Y_1 + \{X_2 Y_2\} + \{X_3 Y_3 Z_3\} + ....$$

where $X_n$ and $Z_n$ are pure in all the coordinates and $Y_n$ is pure in the momenta; therefore $X_n, Y_n$ and $Z_n$ are Hermitian. Then it is stated in [1] that if

(A1.9a) $$\begin{aligned} &Q^j P_k - P_k Q^j = i\hbar \delta_k^j I; \quad P_j P_k = P_k P_j; \quad Q_j Q_k = Q_k Q_j; \\ &q^j \to Q^j; \quad p_k \to P_k \end{aligned}$$

then, where $X$ is any operator function of all the coordinates and all the momenta,

(A1.9b) $$X_{,j} \equiv \frac{\partial X}{\partial Q^j} = \lfloor X, P_j \rfloor; \quad X^{,j} \equiv \frac{\partial X}{\partial P_j} = \lfloor Q^j, X \rfloor$$

where

(A1.9c) $x_n(\underline{q}) \to X_n(\underline{Q}); \quad y_n(\underline{p}) \to Y_n(\underline{P}); \quad x \to X; \quad n = 1, 2, ....$

But it is suggested here that $X$ must have a certain form in order that it can satisfy (A1.9) for certain. A *partial proof* goes as follows:

Put

(A1.10a) $X = X_n = (Q^j)^n \Rightarrow (X)_{,j} = \lfloor X, P_j \rfloor = n(Q^j)^{n-1}; \quad X^{,j} = 0$

Prove by induction or otherwise that

(A1.10b) $\frac{1}{i\hbar}(X_n P_j - P_j X_n) = n(Q^j)^{n-1}$

This proves that $X = X_n$ can be pure polynomial in all the coordinates and that the first part of (A1.9b) is satisfied.

Put

(A1.11a) $X = Y_m = (P_j)^n \Rightarrow X_{,j} = 0; \quad X^{,j} = \frac{1}{i\hbar}(Q^j X - XQ^j) = n(P_j)^{n-1}$

Prove by induction or otherwise that

(A1.11b) $\frac{1}{i\hbar}(Q^j Y_m - Y_m Q^j) = n(P_j)^{n-1}$

This proves that $X = Y_m$ can be pure polynomial in all the momenta and that the second part of (A1.9b) is satisfied.

Put

(A1.12a) $X = \{X_n Y_m\}$

We get ostensibly

(A1.12b)

$$X_{,j} \equiv \frac{\partial X}{\partial Q^j} = \frac{1}{2}(\frac{\partial X_n}{\partial Q^j} Y_m + Y_m \frac{\partial X_n}{\partial Q^j}); \quad \text{Partial Differentiation?}$$

$$= \frac{1}{2}(\lfloor X_n, P_j \rfloor Y_m + Y_m \lfloor X_n, P_j \rfloor); \quad \text{Quantum Mechanics?}$$

$$X^{,k} \equiv \frac{\partial X}{\partial P_k} = \frac{1}{2}(\frac{\partial Y_m}{\partial P_k} X_n + X_n \frac{\partial Y_m}{\partial P_k}); \quad \text{Partial Differentiation?}$$

$$= \frac{1}{2}(\lfloor Q^k, Y_m \rfloor X_n + X_n \lfloor Q^k, Y_m \rfloor); \quad \text{Quantum Mechanics?}$$

To show that (A1.9b) is satisfied, in its entirety, we must have got the partial differentiation right; that is it is assumed here that the product rule for formal partial differentiation applies to Hermitian operators as it does to scalar functions. This is debatable. We have already shown that

(A1.13) $\quad (X_n)_{,j} = \lfloor (X_n), P_j \rfloor; \quad (Y_m)^{,k} = \lfloor Q^k, (Y_m) \rfloor$

We cannot go any further without resolving the queries in (A1.12).

## A2. $H$ Is Quadratic In The $P_j$

Now (A1.3a) is linear in $H$; so the various terms in $H$ will linearly superpose providing that the coefficients are constant. Suppose that

(A2.1) $\quad H \equiv \frac{1}{2}(A^k(\underline{Q})P_k + P_k A^k(\underline{Q})) + B(\underline{Q}) = \{A^k, P_k\} + B \quad$ Hermitian

where $\underline{Q}$ denotes that the $A^k$ and $B$ are pure operator functions of all the coordinate operators and therefore Hermitian. Then

(A2.2) $\quad H^{,l} = A^l \quad$ See (A1.4a)

So that the LHS of (1.5) is equal to

(A2.3a)
$$\begin{aligned}&\frac{1}{i\hbar}(\Theta H - H\Theta)\\ &= \frac{1}{i\hbar}(\Theta\{A^k, P_k\} - \{A^k, P_k\}\Theta); \quad B\Theta = \Theta B\\ &= \frac{1}{2}\lfloor \Theta, P_j \rfloor A^j + \frac{1}{2} A^j \lfloor \Theta, P_j \rfloor; \quad A^k\Theta = \Theta A^k\\ &= \frac{1}{2}(\Theta,_j A^j + A^j \Theta,_j)\end{aligned}$$

which is obviously equal, in this case, to the RHS of (1.5); see (2.2). So $H$ , given by (A2.1), satisfies (A1.5).

Now suppose that we give up the tensor notation and define

(A2.4a) $$H \equiv P^n; \quad n = 1,2,3.... \Rightarrow H' \equiv \frac{\partial H}{\partial P} = nP^{n-1}$$ Hermitian

where

(A2.4b) $$QP - PQ = i\hbar I$$

and I is the unit operator. We also define

(A2.4c) $$\Theta' \equiv \frac{\partial \Theta}{\partial Q} = \frac{1}{i\hbar}(\Theta P - P\Theta); \quad H' = \frac{\partial H}{\partial P} = \frac{1}{i\hbar}(QH - HQ);$$ See (A1.4a)

Therefore

(A2.5a)
$$\begin{aligned}&\frac{1}{i\hbar}(\Theta H - H\Theta) = \frac{1}{i\hbar}(\Theta P^n - P^n\Theta)\\ &= \frac{(\Theta P - P\Theta)}{i\hbar}P^{n-1} + P^{n-1}\frac{(\Theta P - P\Theta)}{i\hbar} + \frac{P\Theta P^{n-1} - P^{n-1}\Theta P}{i\hbar})\\ &= \Theta' P^{n-1} + P^{n-1}\Theta' + \frac{P\Theta P^{n-1} - P^{n-1}\Theta P}{i\hbar}\end{aligned}$$

(A2.5b) $\frac{1}{2}(\Theta' H' + H'\Theta') = \frac{n}{2}(\Theta' P^{n-1} + P^{n-1}\Theta')$

The quantities (2.5a/b) are equal only if

(A2.5c) $n = 2$

and that being so (2.4a) satisfies the appropriate version of (1.5) and

(A2.5d) $\frac{1}{i\hbar}(\Theta H - H\Theta) = \frac{1}{2}(\Theta' H' + H'\Theta')$

Hence $H$ is *generally quadratic* in $P$ ; see (A1.3a/A2.3a).

Now return to the tensor notation with the Einstein convention. Suppose that

(A2.6a) $H = \{G^{jk}(\underline{Q}), P_j P_k\} + P_j F^{jk}(Q) P_k$ Hermitian

where

(A2.6b) $$\begin{array}{ll} g^{jk}(\underline{q}) \to G^{jk}(\underline{Q}); & g^{jk}(\underline{q}) = g^{kj}(\underline{q}); \\ f^{jk}(\underline{q}) \to F^{jk}(\underline{Q}); & f^{jk}(\underline{q}) = f^{kj}(\underline{q}) \end{array}$$

where the $g$ ' s and the $f$ ' s are real free functions of all the coordinates. Now if $X$ is a polynomial operator pure in all the coordinate operators

(A2.7)

$$X_{,jk} = \frac{1}{(i\hbar)^2}(XP_jP_k + P_jP_kX - 2P_jXP_k) \Rightarrow P_jXP_k = \{X, P_jP_k\} - \frac{(i\hbar)^2}{2}X_{,jk}$$

We have

(A2.8a)
$$\begin{aligned}2P_jF^{jk}P_k &= -P_j(P_kF^{jk} - F^{jk}P_k) + (P_kF^{jk} - F^{jk}P_k)P_j \\ &+P_jP_kF^{jk} + F^{jk}P_kP_j \\ &= i\hbar(P_kF^{jk}_{,j} - F^{jk}_{,j}P_k) + P_jP_kF^{jk} + F^{jk}P_kP_j \\ &= -(i\hbar)^2 F^{jk}_{,jk} + P_jP_kF^{jk} + F^{jk}P_kP_j\end{aligned}$$
See (1.4a)

where

(A2.8b) $F^{jk}_{,j,k} \equiv F^{jk}_{,jk}$

because $F^{jk}$ is a pure function of the $\underline{Q}$. It follows that the term $P_jF^{jk}P_k$ (see (A2.6a)) can be subsumed into the term $\{G^{jk}, P_jP_k\}$ and the term $B$ if there is one (see (A2.6a) and (A2.1)). Therefore we may consider only

(A2.8c) $H = \{G^{jk}, P_jP_k\}$ See (A2.6a)

In practice we use, in the main text, only

(A2.9) $n_c = 4$

So, by an appropriate choice of coordinates, we may always make the matrix $G^{ij}$
diagonal; there are only three unique $G^{ij}$ such that $i \neq j$ when $n_c = 4$ whereas when $n_c = 5$ there are ten. Another consequence of the linearity of (A1.5/7), with respect to $H$, is that the argument from (A2.4a) to (A2.6d) may be adapted to prove $H$ is quadratic ($n = 2$) in the case

(A2.10) $H = \{G^{ij}, (P_j)^n\}$; Einstein convention in force

Therefore the most general form of the Hamiltonian operator allowed by the first level of the hierarchy is

(A2.11a) $H = s\{G^{jk}, P_jP_k\} + \{A^k, P\} + B$

where $G^{jk}$, $A^k$ and $B$ are pure operator functions of all the coordinates and therefore Hermitian and $s$ is scalar.

(A2.11b) $$\begin{aligned} &g^{jk}(\underline{q}) \to G^{jk}(\underline{Q}); \quad g^{jk}(\underline{q}) = g^{kj}(\underline{q}); \\ &a(\underline{q}) \to A(\underline{Q}); \quad b(\underline{q}) \to B(\underline{Q}) \end{aligned}$$

# A3. Hamilton's Equations

If the first level in the hierarchy is satisfied the Hamiltonian operator $H$ is quadratic in the momenta $\underline{P}$; let us assume that this is so. The classical Hamilton's equations are

(A3.1a)
$$\dot{p}_j = -\frac{\partial h}{\partial q^j}; \quad \dot{q}^k = \frac{\partial h}{\partial p_k} \Rightarrow \dot{p}_l = -p_j p_k g^{jk}_{,l} - p_r f^r_{,l} - v_{,l}; \quad \dot{q}^l = 2 p_l g^{jl} + f^l$$

where

(A3.1b) $$h = p_j p_k g^{jk} + p_r f^r + v$$

is the scalar Hamiltonian, $q^k$ coordinates and $p_j$ the momenta and dot denotes differentiation with respect to time; the coefficients $g^{jk} = g^{kj}$, $f^r$ and v are pure functions of all the scalar coordinates. The quantum mechanical version of (A3.1a) is

(A3.2a) $$\dot{P}_l = -\{P_j P_k, G^{jk}_{,l}\} - \{P_r, F^r_{,l}\} - V_{,l}; \quad \dot{Q}^l = 2\{P_l, G^{jl}\} + F^l$$

where

(A3.2b) $$H = \{P_j P_k, G^{jk}\} + \{P_r, F^r\} + V$$

and $G^{jk} = G^{kj}, F^r$ and $V$ are pure polynomials in all the coordinates. It is stated in [2] that things can be arranged so that Hamilton's equations are true in both quantum mechanics and classical mechanics; this is not generally true as the above argument shows. It is only true when the space is flat, $f^r = 0$ and the coordinates are Cartesian.

# Appendix B - If The First Two Levels Of The Hierarchy Are Satisfied - The Theta Equation And GTE

## B1. Introduction

It is desirable that $H$ and $\Theta$ should satisfy as many constraints as possible, consecutively, beginning with constraint 1. We have already proved (in Appendix A) that if $H$ and $\theta \to \Theta$ satisfy constraint 1 then $H$ is quadratic in the $P_j,\ j = 1,2,....n_c$

(B1.1a) $H = \mathrm{K}\left\{G^{uv}, P_u P_v\right\} + \left\{F^j, P_j\right\} + V;$ $\mathrm{K}$ scalar; $\left\{\mathrm{A,B}\right\} = \frac{1}{2}(AB + BA)$

where the operators $G^{uv}$, $F^j$ and $V$ pure in the coordinate operators $Q^k$ . We are only interested in the gravitational case where

(B1.1b) $F^j = O$ and $V = O$

Therefore

(B1.1c) $H = \mathrm{K}\left\{G^{uv}, P_u P_v\right\}$

where in the coordinate representation

(B1.2a) $P_j = -i\hbar \dfrac{\partial}{\partial q^j}; \quad \mathrm{Q}^{\mathrm{k}} = q^k I; \quad i\hbar\delta^j_k I = -i\hbar\left( q^j \dfrac{\partial}{\partial q^k} - \dfrac{\partial}{\partial q^k} q^j \right);$

(B1.2b) $G^{uv}(\underline{Q}) \equiv g^{uv}(\underline{q})I = G^{vu}(\underline{Q}); \quad F^j(\underline{Q}) \equiv f^j(\underline{q})I; \quad V(\underline{Q}) \equiv v(\underline{q})I$

where $\underline{q} \to \underline{Q}$ denotes the aggregate of the $q^j \to Q^j$. In view of the first sentence of this Introduction we consider, at the least, constraint 2 as well as constraint 1 should be satisfied; we expect $\theta \to \Theta$ to be restricted thereby.

## B2. Constraints 1 And 2

With the notation

(B2.1a) $\lfloor A, B \rfloor \equiv \dfrac{1}{i\hbar}(AB - BA); \quad \{A, B, ....\} \equiv \dfrac{1}{n!}\sum_{perm}(A, B, ....);$

(B2.1b) $A^{:j} \equiv \lfloor Q^j, A \rfloor = \dfrac{\partial A}{\partial P_j}; \quad A_{,j} \equiv \lfloor A, P_j \rfloor = \dfrac{\partial A}{\partial Q^j}; \quad j = 1, 2, ...n_c$

(B2.1c)

$$\frac{dA}{dt} \equiv -\frac{1}{i\hbar}(HA - AH) = -\lfloor H, A \rfloor; \quad \frac{d^2 A}{dt^2} \equiv \lfloor H, \lfloor H, A \rfloor \rfloor \equiv \lfloor H, H, A \rfloor$$

Constraint 1 is ( $()^{:j} \equiv ()^{,j}$ )

(B2.2a) $Z_1 \equiv \left\{ H^{:j}, \Theta_{,j} \right\} = -\lfloor H, \Theta \rfloor = \lfloor \Theta, H \rfloor;$ Einstein convention in force

Constraint 2 is

(B2.2b) $Z_2 \equiv \left\{ \lfloor H, H^{:j} \rfloor, \Theta_{,j} \right\} + \left\{ H^{:j}, H^{:k}, \Theta_{,j,k} \right\} = \lfloor H, H, \Theta \rfloor$

By combining this with (B2.2a) we can remove all explicit reference to the $H^{:j}$ and to $H$. We obtain thereby an operator equation involving only the derivations of $\Theta$. In the coordinate representation this reduces to a fourth order PDE satisfied by $\theta$. The PDE contains no reference to either the $f^u$ or $v$. It allows us, in principle, to calculate functions $\theta(\underline{q})$ that satisfy both constraints 1 and 2 given the functions $g^{uv}$.

Differentiate (B2.2a) with respect $t$ to produce

(B2.3) $$-\lfloor H, Z_1 \rfloor = -\lfloor H, \{H^{:j}, \Theta_{,j}\} \rfloor = -\lfloor H, \lfloor H, \Theta \rfloor \rfloor = -\lfloor H, H, \Theta \rfloor$$

Add (B2.3) from (B2.2b) to get

(B2.4)
$$Z_2 - \lfloor H, Z_1 \rfloor \Rightarrow \{\lfloor H, H^{:j} \rfloor, \Theta_{,j}\} - \lfloor H, \{H^{:j}, \Theta_{,j}\} \rfloor + \{H^{:j}, H^{:k}, \Theta_{,j,k}\} = O$$

Now we have

(B2.5a) $$\Theta_{,j} \equiv \frac{\partial \Theta}{\partial Q^j} = \frac{1}{i\hbar}(\Theta P_j - P_j \Theta) = \lfloor \Theta, P_j \rfloor$$

(B2.5b) $$H^{:j} \equiv \frac{\partial H}{\partial P_j} = \frac{1}{i\hbar}(Q^j H - H Q^j) = \lfloor Q^j, H \rfloor = \dot{Q}^j; \quad \text{See (B2.1)}$$

(B2.6) $$\begin{aligned} &\{\lfloor H, H^{:j} \rfloor, \Theta_{,j}\} - \lfloor H, \{H^{:j}, \Theta_{,j}\} \rfloor \\ &= \{\ddot{Q}^j, \Theta_{,j}\} - \{\dot{H}^j, \Theta_{,j}\} = \{\ddot{Q}^j, \Theta_{,j}\} - \{\ddot{Q}^j, \Theta_{,j}\} = O; \quad \dot{H} = O \end{aligned}$$

The remaining term is

(B2.7a)
$$\{H^{:j}, H^{:k}, \Theta_{,j,k}\} = \{H^{:j}, \{H^{:k}, \Theta_{,j,k}\}\} = O$$
$$\Rightarrow \frac{1}{12}(H^{:j} H^{:k} \Theta_{,j,k} + \Theta_{,j,k} H^{:j} H^{:k} - 2H^{:j} \Theta_{,j,k} H^{:k}) = O; \quad \Theta_{,j,k} = \Theta_{,k,j}$$

written in full. An alternative to this is (B2.7b) $\frac{(i\hbar)^2}{12}\left[H^{:j},\left[H^{:k},\Theta_{,j,k}\right]\right]=O; \quad \Theta_{,j,k}=\Theta_{,k,j}$

Because

(B2.8) $\lfloor H^{:l},A\rfloor = 2\mathrm{K}G^{ul}A_{,u}; \quad AQ^u = Q^uA;$ see (B2.1a)

the result (B2.7b) is identical to

(B2.9a) $(i\hbar)^2\,\frac{\mathrm{K}}{6}\left[H^{:j},G^{uk}\Theta_{,j,k,u}\right]=O$

giving, upon further application of (B2.8),

(B2.9b) $-\frac{\hbar^2\mathrm{K}^2}{3}G^{vj}(G^{uk}\Theta_{,j,k,u})_{,v}=O; \quad j,k,u,v=1,2,...n_c$ Operator Theta Equation

The numerical factor $-\hbar^2\mathrm{K}^2/3$ can, of course, be cancelled; we retain this factor at (B2.9b) because the operator on the LHS is the imbalance across (B2.2b). In the coordinate representation

(B2.10a) $G^{uv}\equiv g^{uv}(\underline{q})I; \quad g^{uv}=g^{vu}; \quad \Theta=\theta(\underline{q})I$

and (B2.9b) reduces to the PDE

(B2.10b) $g^{vj}(g^{uk}\theta_{,jku})_{,v}=0; \quad j,k,u,v=1,2,...n_c$; Scalar Theta Equation

Notice that the theta equation does not contain the functions $f^j$ and $v$. Further the theta equation is not a classical approximation; it is a purely QM result

That $\theta$ satisfies the PDE (B2.10b) raises an immediate issue: Quadratic operator $H$ derives from constraint 1 on the assumption that $\theta$ is arbitrary; but it cannot be truly arbitrary if it satisfies (B2.10b). At most it is the general solution of given a particular $n_c$-space C. So is the quadratic form of the operator $H$ valid? Yes! Solutions of (B2.10b). must be subject to complicated boundary conditions; thus $\theta$ is sufficiently arbitrary for the quadratic solution of constraint 1 to follow.

The scalar theta equation (B2.10b) can be regarded as a field equation for theta (subject to possible modification by higher constraints at levels 3 and above). Because the $g^{lm}$ the $f^{l}$ and $v$ inform the Hamiltonian they are all candidates for $\theta$. We thus have three versions of (B2.10b) that are putative field equations for the $g^{lm}$ the $f^{l}$ and $v$:

(B2.11) $g^{vj}(g^{uk}g^{lm}{}_{,jku})_{,v} = 0; \quad g^{vj}(g^{uk}f^{l}{}_{,jku})_{,v} = 0; \quad g^{vj}(g^{uk}v_{,jku})_{,v} = 0$

Recall that, according to the quantization axioms, these equations are true only in a flat space using flat (e.g., Cartesian) coordinates.

We support the (unproven) conjecture the if the first two levels of the hierarchy of constraints is satisfied then that is all that is required to discuss conventional CM.

# Appendix C- Results For Calculations (13.4) To (13.12)-Taken From The Output Of Maple 16

The original equations are

$$A := a0 + a1\,u + a2\,u^2 + a3\,u^3 + a4\,u^4 + a5\,u^5$$

$$p := p0 + p1\,u$$

$$\rho := \rho 0 + \rho 1\,u$$

$$o4 := \left(a0 + a1\,u + a2\,u^2 + a3\,u^3 + a4\,u^4 + a5\,u^5\right)^2 \left(2\,a2 + 6\,a3\,u + 12\,a4\,u^2 + 20\,a5\,u^3\right)^2 + 4\,k\left(a1 + 2\,a2\,u + 3\,a3\,u^2 + 4\,a4\,u^3 + 5\,a5\,u^4\right)^2 + 4\left(a1 + 2\,a2\,u + 3\,a3\,u^2 + 4\,a4\,u^3 + 5\,a5\,u^4\right)^4 - 5\left(a0 + a1\,u + a2\,u^2 + a3\,u^3 + a4\,u^4 + a5\,u^5\right)\left(a1 + 2\,a2\,u + 3\,a3\,u^2 + 4\,a4\,u^3 + 5\,a5\,u^4\right)^2 \left(2\,a2 + 6\,a3\,u + 12\,a4\,u^2 + 20\,a5\,u^3\right) + \left(a0 + a1\,u + a2\,u^2 + a3\,u^3 + a4\,u^4 + a5\,u^5\right)^2 \left(a1 + 2\,a2\,u + 3\,a3\,u^2 + 4\,a4\,u^3 + 5\,a5\,u^4\right)\left(6\,a3 + 24\,a4\,u + 60\,a5\,u^2\right) + \left(a0 + a1\,u + a2\,u^2 + a3\,u^3 + a4\,u^4 + a5\,u^5\right)^3 (24\,a4 + 120\,a5\,u)$$

$$den := k + \left(a1 + 2\,a2\,u + 3\,a3\,u^2 + 4\,a4\,u^3 + 5\,a5\,u^4\right)^2 - \frac{1}{3}\left(\chi\,c^2\,(\rho 0 + \rho 1\,u) + \Lambda\right)\left(a0 + a1\,u + a2\,u^2 + a3\,u^3 + a4\,u^4 + a5\,u^5\right)^2$$

$$pres := k + \left(a1 + 2\,a2\,u + 3\,a3\,u^2 + 4\,a4\,u^3 + 5\,a5\,u^4\right)^2 + 2\left(a0 + a1\,u + a2\,u^2 + a3\,u^3 + a4\,u^4 + a5\,u^5\right)\left(2\,a2 + 6\,a3\,u + 12\,a4\,u^2 + 20\,a5\,u^3\right) + \left(\chi\,(p0 + p1\,u) - \Lambda\right)\left(a0 + a1\,u + a2\,u^2 + a3\,u^3 + a4\,u^4 + a5\,u^5\right)^2$$

Here and now

$$u := 0$$

$$o10 := 90\,a0^2\,a1\,a3 + 44\,a0^2\,a2^2 + 72\,a0^3\,a4 + 4\,k\,a1^2 - 4\,a1^4 + 8\,k^2 - 14\,a0\,a1^2\,a2 - 8\,a0\,a2\,k$$

$$o40 := 4\,a0^2\,a2^2 + 4\,k\,a1^2 + 4\,a1^4 - 10\,a0\,a1^2\,a2 + 6\,a0^2\,a1\,a3 + 24\,a0^3\,a4$$

$$den0 := k + a1^2 - \frac{1}{3}\left(\chi\,c^2\,\rho0 + \Lambda\right)a0^2$$

$$pres0 := k + a1^2 + 4\,a0\,a2 + \left(\chi\,p0 - \Lambda\right)a0^2$$

$$a0 := 1$$

$$a1 := \frac{H}{c}$$

The rest of the formulae

$$\left\{a2 = -\frac{1}{12}\,\chi\,c^2\,\rho0 + \frac{1}{6}\,\Lambda - \frac{1}{4}\,\chi\,p0,\ a3 = \frac{1}{108}\,\frac{1}{H\,c}\left(15\,c^2\,H^2\,\chi\,\rho0 - 2\,\rho0^2\,\chi^2\,c^6 - c^4\,\Lambda\,\rho0\,\chi - 3\,c^4\,\chi^2\,\rho0\,p0 - 2\,c^2\,\Lambda^2 + 3\,c^2\,\Lambda\,\chi\,p0 - 3\,c^2\,\chi^2\,p0^2 + 6\,H^2\,\Lambda + 9\,H^2\,\chi\,p0\right),\ a4 = -\frac{1}{288}\,\frac{1}{c^2}\left(\chi\left(36\,c^2\,H^2\,\rho0 - \rho0^2\,\chi\,c^6 - 2\,c^4\,\Lambda\,\rho0 - 2\,c^2\,\Lambda\,p0 + c^2\,\chi\,p0^2 + 36\,H^2\,p0\right)\right),\ k = -\frac{1}{3}\,\frac{3\,H^2 - \chi\,c^4\,\rho0 - \Lambda\,c^2}{c^2}\right\}$$

Results for an empty model universe

$$\rho0 := -\frac{\Lambda}{\chi\,c^2}$$

$$p0 := \frac{\Lambda}{\chi}$$

$$\left\{a2 = 0, a3 = 0, a4 = 0, k = -\frac{H^2}{c^2}\right\}$$

Numerical results (13.5)- Assumptions

$$H := 2.055000000\,10^{-18}$$

$$c := 2.997900000\,10^{8}$$

$$aa1 := 6.854798359\,10^{-27}$$

$$A := a0 + a1\,u + a2\,u^2 + a3\,u^3 + a4\,u^4 + a5\,u^5$$

$$p := p0 + p1\,u$$

$$\rho := \rho0 + \rho1\,u$$

$$u := 0$$

$$o10 := 90\,a0^2\,a1\,a3 + 44\,a0^2\,a2^2 + 72\,a0^3\,a4 + 4\,k\,a1^2 - 4\,a1^4 + 8\,k^2 - 14\,a0\,a1^2\,a2 - 8\,a0\,a2\,k$$

$$o40 := 4\,a0^2\,a2^2 + 4\,k\,a1^2 + 4\,a1^4 - 10\,a0\,a1^2\,a2 + 6\,a0^2\,a1\,a3 + 24\,a0^3\,a4$$

$$den0 := k + a1^2 - \frac{1}{3}\left(\chi\,c^2\,\rho0 + \Lambda\right)a0^2$$

$$pres0 := k + a1^2 + 4\,a0\,a2 + \left(\chi\,p0 - \Lambda\right)a0^2$$

$$k := 0$$

$$\rho0 := 8.000000000\,10^{-27}$$

$$p0 := 0$$

$$a0 := 1$$

$$a1 := \frac{H}{c}$$

$$\chi := 2.076100000\,10^{-43}$$

Results for (13.5)

$$\{\Lambda = -8.305220790\,10^{-54}, a2 = -1.382337033\,10^{-53}, a3 = 8.024424778\,10^{-80}, a4 = -8.079849533\,10^{-106}\}$$

Numerical results for (13.6)- Assumptions

$$A := a0 + a1\,u + a2\,u^2 + a3\,u^3 + a4\,u^4 + a5\,u^5$$

$$p := p0 + p1\,u$$

$$\rho := \rho0 + \rho1\,u$$

$$u := 0$$

$$o10 := 90\,a0^2\,a1\,a3 + 44\,a0^2\,a2^2 + 72\,a0^3\,a4 + 4\,k\,a1^2 - 4\,a1^4 + 8\,k^2 - 14\,a0\,a1^2\,a2 - 8\,a0\,a2\,k$$

$$o40 := 4\,a0^2\,a2^2 + 4\,k\,a1^2 + 4\,a1^4 - 10\,a0\,a1^2\,a2 + 6\,a0^2\,a1\,a3 + 24\,a0^3\,a4$$

$$den0 := k + a1^2 - \frac{1}{3}\left(\chi\,c^2\,\rho0 + \Lambda\right)a0^2$$

$$pres0 := k + a1^2 + 4\,a0\,a2 + \left(\chi\,p0 - \Lambda\right)a0^2$$

$$k := 1.000000000\,10^{-54}$$

$$\rho0 := 8.000000000\,10^{-27}$$

$$p0 := 0$$

$$a0 := 1$$

$$a1 := \frac{H}{c}$$

Results for (13.6)

$$\{\Lambda = -5.305220790\ 10^{-54},\ a2 = -1.332337033\ 10^{-53},\ a3 = 8.089213313\ 10^{-80},\ a4 = -8.048751616\ 10^{-106}\}$$

Numerical results for (13.9)

Solutions of quartic

$$2.500253192\ 10^{26},\ -2.410983337\ 10^{25} + 2.201628897\ 10^{26}\ \mathrm{I},\ -1.013029447\ 10^{26},\ -2.410983337\ 10^{25} - 2.201628897\ 10^{26}\ \mathrm{I}$$

$$age := 2.642936097\ 10^{10}$$

Numerical Results for (13.11)- Assumptions

$$A := a0 + a1\,u + a2\,u^2 + a3\,u^3 + a4\,u^4 + a5\,u^5$$

$$p := p0 + p1\,u$$

$$\rho := \rho 0 + \rho 1\,u$$

$$u := 0$$

$$o10 := 90\,a0^2\,a1\,a3 + 44\,a0^2\,a2^2 + 72\,a0^3\,a4 + 4\,k\,a1^2 - 4\,a1^4 + 8\,k^2 - 14\,a0\,a1^2\,a2 - 8\,a0\,a2\,k$$

$$o40 := 4\,a0^2\,a2^2 + 4\,k\,a1^2 + 4\,a1^4 - 10\,a0\,a1^2\,a2 + 6\,a0^2\,a1\,a3 + 24\,a0^3\,a4$$

$$den0 := k + a1^2 - \frac{1}{3}\left(\chi\,c^2\,\rho 0 + \Lambda\right)a0^2$$

$$pres0 := k + a1^2 + 4\,a0\,a2 + \left(\chi\,p0 - \Lambda\right)a0^2$$

$$k := 1.000000000\ 10^{-54}$$

$$p0 := 0$$

$$a0 := 1$$

Value of $a1$

$$aa1 := 6.854798359\ 10^{-27}$$

Age of the Universe

$$y := 1.296038477\ 10^{26}$$

(13.10)- Extra equation

$$AA := 1.888408242 + 1.679715734\ 10^{52}\ a2 + 2.176976222\ 10^{78}\ a3 + 2.821444947\ 10^{104}\ a4$$

Results- Only take $\Lambda$ negative and $\rho 0$ positive

$$\{\Lambda = 8.079505311\ 10^{-52}, a2 = 1.899905676\ 10^{-52}, a3 = -2.554391430\ 10^{-78}, a4 = 1.705295387\ 10^{-105}, \rho 0 = -3.558575675\ 10^{-26}\}, \{\Lambda = -2.077396280\ 10^{-52}, a2 = -6.393197215\ 10^{-53}, a3 = -9.633534651\ 10^{-80}, a4 = -2.143628883\ 10^{-105}, \rho 0 = 1.884930148\ 10^{-26}\}$$

# Appendix D- The K Equation Revisited

## D1. Geodesic, Canonical And Cartesian Coordinates

The letter [7] raises certain questions; this appendix is an effort answer them. A derivation of the K equation is also given .

The Cristoffel symbols are defined a follows (suffices run from 1 to $n_c$ in this case):

(D1.1a) $[ij,k] \equiv \frac{1}{2}\left(g_{ik,j} + g_{jk,i} - g_{ij,k}\right)$; first kind [1, p. 26]

(D1.1b) $\Gamma^l_{ij} \equiv g^{lk}[ij,k]$ ; second kind

(D1.1c) $\text{Suffices} = [1, n_c]$

where $()_{,j}$ denotes *partial differentiation* with respect to the $j^{th}$ coordinate, $g_{ij}$ is an element of the covariant fundamental tensor and $g^{lk}$ is an element of the contravariant fundamental tensor. The fundamental tensor is transparent to covariant differentiation [1]

(D1.2) $$g_{jk;l} = 0 \text{ and } (g^{jk})_{;l} = 0$$

were $()_{;l}$ denotes *covariant differentiation* with respect to $l^{th}$ coordinate. Covariant differentiation of a product follows the usual rule for partial differentiation [1]. The connection between covariant fundamental tensor $g_{ij}$ and the contravariant fundamental tensor $g^{uv}$ is

(D1.3) $$g^{uv} g_{vw} = \delta^u_w$$ Einstein convention is in force where $\delta^u_u = 1$ and $\delta^{\text{u}}_{\text{v}} = 0$ for $u \neq v$ [1].

A coordinate is *geodesic* at pole *P* if

(D1.4) $$[ij,k] = 0$$

at pole *P*. It follows that the Christoffel symbols of the first and second kinds vanish at the pole; therefore the first covaritant derivative is equal to the first partial derivative at *P* ; see (D1.1a/b) [1].

Coordinates that satisfy, at a pole *P*,

(D1.5a) $$\Gamma^a_{bc,d} + \Gamma^a_{cd,b} + \Gamma^a_{db,c} = 0;$$ ; constrains the $g_{ab,cd}$; $n_c = 4$

are said to be *canonical*. Does the condition (D1.5a) define a valid set of coordinates? To decide that we must investigate whether or not the conditions (D1.5a) could constrain the curvature of the space (defined by the Riemann-Christoffel tensor $R^u_{vwx}$) at *P*. Because $\Gamma^a_{bc}$ is symmetrical in the suffices $b, c$ it follows from (D1.5a) that

(D1.5b) $$\Gamma^a_{cb,d} + \Gamma^a_{dc,b} + \Gamma^a_{bd,c} = 0$$

There are thus

(D1.6) $$n_c\left[n_c+n_c(n_c-1)+\frac{n_c(n_c-1)(n_c-2)}{3!}\right]=n_c^2(n_c^2+3n_c+2)/6$$

unique conditions which constrain the

(D1.7) $$n_c^2(n_c+1)^2/4$$

second derivatives of the $g_{uv}$ at *P*. It follows that there are still

(D1.8) $$n_c^2(n_c+1)^2/4-n_c^2(n_c^2+3n_c+2)/6=n_c^2(n_c^2-1)/12$$

degrees of freedom. This is also the number of unique, independent elements of the curvature tensor $R^u_{vwx}$ [1]. So the conditions (D1.5a/b) do not constrain the curvature at *P*.

The counting of the unique conditions at (D1.6) goes as follows: $a$ takes $n_c$ values independently of $b,c$ and $d$. So, in the square brackets on the LHS of (1.6), we have the contribution as $b,c$ and $d$ vary. There are $n_c$ cases for which $b=c=d$. There are $n_c(n_c-1)$ cases for which two of the suffices $b,c,d$ are equal. Because, permutation does not increase the number of conditions when the suffices $b,c,d$ all differ, there are in total $n_c(n_c-1)(n_c-2)/3!$ cases for which none of these suffices are equal.

Conditions (D1.5a) constrain the second derivatives of the fundamental tensor at *P*. But we have yet to constrain the $g_{uv}$ themselves. The Schrodinger quantisation rules appear to require that definitions be couched in terms of flat, Cartesian coordinates $\underline{q}$. A curved manifold $C'$ can be made flat at *P*, and approximately flat in a neighbourhood of *P*, by requiring that *P* is the pole of local *Cartesian geodesic* (CG) coordinates. Thus the requirements of the Schrodinger definitions are met, in a neighbourhood of a point *P*, if the coordinates are chosen to be CG with pole *P*. There may be convenience of calculation if, in addition, the coordinates are chosen to be *canonical* at *P*; the coordinates are then said to be CCG (Cartesian canonical geodesic) with pole *P*.

## D2. Questions About The Gravitational Theta Equation

The gravitational theta equation (GTE) can be written

(D2.1) $$g^{vj}(g^{uk}g^{lm}_{,jku})_{,v}=0; \quad ()_{,jkuv}\equiv()_{,j,k,u,v} \text{ by convention;}$$
$$j,k,l,m,u,v=1,2,...n_c; \quad n_p=1; \quad n_c=4$$

where the coordinates and the space are defined as flat. We suppose that, in a curved Riemannian manifold $C'$, the GTE holds only at the pole *P* of CG coordinates. Solutions of the GTE, as it stands, then approximate the manifold in the neighbourhood of *P*. The question arises: What tensor equation reduces to (D2.1) at the pole $P$ in the coordinates chosen ? Such a tensor equation would be valid in any coordinate system at all points of the manifold. Clive Kilmister derived a tensor equation that purports to satisfy this condition; but his argument is based on a version of the GTE that depends upon $g_{ab,cde}$. Is this version valid? He takes the theta equation

(D2.2) $$g^{vj}(g^{uk}\theta_{,jku})_{,v}=0$$

and simply defines

(D2.3) $$\theta\equiv g_{ab}$$

to get a GTE

(D2.4) $$g^{vj}(g^{uk}g_{ab,jku})_{,v}=0; \quad \text{Kilmister GTE}$$

whereas it is usual to define the GTE using the coefficients of the quadratic terms of the Hamiltonian

(D2.5) $$\theta\equiv g^{lm}; \text{ see (D2.1)}$$

We need to explore the relation between the LHSs of (D2.1) and (D2.4) to answer this question. Note that, in geodesic coordinates at the pole *P*, (D2.1) simplifies to

(D2.6) $$g^{vj}(g^{uk}g^{lm}_{,jku})_{,v} = g^{vj}(g^{uk}g^{lm}_{,jku})_{;v} = g^{vj}g^{uk}g^{lm}_{,jkuv} = 0$$ ; canonical GTE in geodesic coordinates

Whereas, in the same coordinates, (D2.4) simplifies to

(D2.7) $g^{vj}g^{uk}g_{ab,jkuv} = 0$; Kilmister's form of the GTE in geodesic coordinates

Now choose the metric (2.7) (main text). We have

(D2.8a) $g_{aa} \approx -1 + 2U; \quad g^{ll} \approx -1 - 2U; \quad |U| << 1; \quad a,l = [1,3]$

(D2.8b) $g_{aa,jkuv} \approx 2U_{,jkuv}; \quad g^{ll}_{,jkuvv} \approx -2U_{,jkuv}; \quad |U| << 1; \quad a,l = [1,3]$

(D2.8c) $n_p = 1; \quad n_c = 4$

where

(D2.9) $U_{,4} = 0$; $U$ depends upon the first three coordinates only

The canonical GTE (D2.6) becomes

(D2.10) $g^{vj}g^{uk}g^{ll}_{,jkuv} \approx -2g^{vj}g^{uk}U_{,jkuv} = 0$

The Kilmister form of the GTE (D2.7) becomes

(D2.11) $g^{vj}g^{uk}g_{ab,jkuv} \approx 2g^{vj}g^{uk}U_{,jkuv} = 0$

So the two forms of the GTE are equivalent.

But what of the condition attached to (D2.8a/b)? This surely means that we have proved equivalence only for *weak gravity*. On the contrary; however strong the gravity we can always choose the pole *P* of the CG coordinates so that

(D2.12) $U = 0 \Rightarrow U_{,j} = 0; \quad U_{jk} = 0; \quad U_{jkl} = 0; \quad U_{jklm} = 0$

at *P*; hence we can write exact equalites at (D2.10/11).

## D3. Lemmas

We begin to derive the K equation with certain lemmas. What follows is an identity

(D3.1a) $g_{ab,c} = g_{pb}\Gamma^{p}_{ac} + g_{ap}\Gamma^{p}_{bc}$; [1, equ. (20.4) et seq., p.27] Identity

So we have in geodesic coordinates pole *P*,

(D3.2) $g_{ab,cd} = g_{pb}\Gamma^{p}_{ac,d} + g_{ap}\Gamma^{p}_{bc,d}$; in geodesic coordinates pole *P*

Now, also at *P* in geodesic coordinates, [1, p. 49 et seq.]

(D3.3) $R^{a}_{bcd} = \Gamma^{a}_{bd,c} - \Gamma^{a}_{bc,d}$; Riemann-Christoffel or curvature tensor [1]

where, from (D1.1b/D1.5a),

(D3.4) $-\Gamma^{a}_{bc,d} = \Gamma^{a}_{cd,b} + \Gamma^{a}_{db,c}$; CCG coordinates pole *P*

so that (D3.3) becomes

(D3.5) $R^{a}_{bcd} = 2\Gamma^{a}_{bd,c} + \Gamma^{a}_{cd,b}$; CCG coordinates pole *P*

Interchange $b$ and $c$

(D3.6) $R^a_{cbd} = \Gamma^a_{bd,c} + 2\Gamma^a_{cd,b}$; CCG coordinates pole *P*

These two (D3.5/6) give

(D3.7) $3\Gamma^a_{cd,b} = 2R^a_{cbd} - R^a_{bcd}$; CCG coordinates pole *P*

But

(D3.8) $-R^a_{bcd} = R^a_{cdb} + R^a_{dbc} = -R^a_{cbd} + R^a_{dbc}$; [1, p. 50]

Hence,

(D3.9a) $\Gamma^a_{cd,b} = \frac{1}{3}(R^a_{cbd} + R^a_{dbc})$; CCG coordinates pole *P*

Therefore

(D3.9b) $$\begin{aligned}[ij,m]_{,k} &= (g_{lm}\Gamma^l_{ij})_{,k} = g_{lm}\Gamma^l_{ij,k} \\ &= \tfrac{1}{3}(R_{mikj} + R_{mjki})\end{aligned}$$; CCG coordinates pole *P*

So, using (D3.2/9a) and the properties of $R_{acdb}$ [1, p. 51]

(D3.10a) $$\begin{aligned}g_{ab,cd} &= g_{pb}\Gamma^p_{ac,d} + g_{ap}\Gamma^p_{bc,d} \\ &= \tfrac{1}{3}[g_{pb}(R^p_{adc} + R^p_{cda}) + g_{pa}(R^p_{bdc} + R^p_{cdb})] \\ &= \tfrac{1}{3}(R_{badc} + R_{bcda} + R_{abdc} + R_{acdb}) \\ &= \tfrac{1}{3}(R_{cbad} + R_{cabd})\end{aligned}$$; CCG coordinates pole *P*

because $R_{ijkl}$ is skew symmetric in $ij$ and $kl$. Therefore, because by definition

(D3.10b) $g^{cd}R_{cbad} = R^d_{bad} = R_{ba} = R_{ab}$,

(D3.10c) $g^{cd}g_{ab,cd} = \frac{1}{3}g^{cd}(R_{cbad} + R_{cabd}) = \frac{2}{3}R_{ab}$; CCG coordinates pole *P*

Further it follows from (D3.10a) that

$$\begin{aligned} g^{aj}_{,de} &= -g^{jk}g^{ab}g_{bk,de} \\ &= -\tfrac{1}{3}g^{jk}g^{ab}(R_{dkbe} + R_{dbke}) \\ &= \tfrac{1}{3}g^{ab}(R^{j}_{dbe} + R^{j}_{ebd}) \end{aligned} \quad \text{(D3.11); CCG coordinates pole } P$$

We have now the formulae necessary to express the first derivatives of the Christoffel symbols and the second derivatives of the fundamental tensor in terms of the curvature tensor. Most of them were derived by Clive Kilmister.

## D.4 The K Equation

We need, in the course of the final argument, to express $g^{ef}R_{ab,ef}$ in terms of tensors evaluated in suitable coordinates. By definition

(D4.1) $R_{ab;e} \equiv R_{ab,e} - \Gamma^{l}_{ae}R_{lb} - \Gamma^{l}_{be}R_{al}$; [1, p. 34] Identity/ definition

where ';' denotes covariant differentiation. Differentiating covariantly again, but expressing the result using geodesic coordinates with pole *P*,

(D4.2a) $R_{ab;ef} = R_{ab,ef} - \mu_{abef}$; in geodesic coordinates pole *P*

where, in the same coordinates,

(D4.2b) $\mu_{abef} \equiv \Gamma^{l}_{ae,f}R_{lb} + \Gamma^{l}_{be,f}R_{al}$; in geodesic coordinates pole *P*

Therefore

(D4.3a) $g^{ef}R_{ab,ef} = g^{ef}R_{ab;ef} + M_{ab}$; in geodesic coordinates pole *P*

where

(D4.3b) $M_{ab} \equiv g^{ef}\mu_{abef}$ ; in geodesic coordinates pole *P*

From (D3.9a/4.3b)

(D4.4)
$$\begin{aligned} M_{ab} &= g^{ef}\ \Gamma^{l}_{ae,f}R_{lb} + \Gamma^{l}_{be,f}R_{al} \\ &= \tfrac{1}{3}g^{ef}\left[\ R^{l}_{afe} + R^{l}_{efa}\ \ R_{lb} +\ R^{l}_{bfe} + R^{l}_{efb}\ \ R_{al}\right] \\ &= \tfrac{1}{3}g^{ef}\left[R^{l}_{efa}R_{lb} + R^{l}_{efb}R_{al}\right] \end{aligned}$$
; CCG coordinates pole *P*

Now

(D4.5)
$$g^{ef}\left[R^{l}_{efa}R_{lb} + R^{l}_{efb}R_{al}\right] = g^{ef}g^{lr}\left[R_{refa}R_{lb} + R_{refb}R_{la}\right] = g^{lr}\left[R_{ra}R_{lb} + R_{rb}R_{la}\right]$$

Because $l,e$ and $r,f$ are dummy

(D4.6) $M_{ab} = \frac{1}{3}g^{ef}\left[R_{fa}R_{eb} + R_{fb}R_{ea}\right] = \frac{2}{3}g^{ef}R_{ea}R_{fb}$ ; CCG coordinates pole *P*

$M_{ab}$ is obviously a tensor. So is

(D4.7) $K_{ab} \equiv g^{ef}R_{ab,ef} = g^{ef}R_{ab;ef} + M_{ab}$; See (D4.3a) ; CCG coordinates pole *P*

Now we have shown that (see Section 6)

(D4.8)

$$R_{ab} = R_{ab} \approx -\Delta_{ab}\nabla^2 U; \quad \left|U\right| << 1; \quad a,b = 1,2,3,4; \quad n_p = 1; \quad n_c = 4$$
$$g^{ef}R_{ab;ef} \approx -\Delta_{ab}\nabla^2(\nabla^2 U)$$

Therefore

(D4.9a) $K_{ab} \equiv g^{ef}(R_{ab;ef} + \frac{2}{3} R_{ae} R_{fb})$

(D4.9b) $K_{aa} \approx (-1+U) \; -\nabla^2(\nabla^2 U) + \frac{2}{3}(\nabla^2 U)^2 \;\; \approx \nabla^2(\nabla^2 U); \quad a = \left[1,3\right]$

(D4.10c) $K_{ab} = 0; \quad a \neq b; \quad 1 >> \left|U\right|$

However strong the gravity we can always choose the pole *P* of the CG coordinates so that (D2.12) is satisfied. Hence

(D4.12) $K_{ab} = 0$

# 14. Conclusions

There is a formulation of Quantum Mechanics (QM) which relies not on energy equations (Schrodinger) or path integrals with integrands of Lagrangians (Feynman) but on something which is *inevitable*. Namely an infinite hierarchy *identities* associated with operators $\Theta$.

Each operator $\Theta$ depends only upon each of the (Cartesian) coordinate operators in an flat Riemannian $n_c$ -space *C* and goes to define the 'system'.

If the first (lowest level) operator identity is satisfied, in one these hierarchies, the Hamiltonian operator is quadratic in the operators that are usually regarded as the (Cartesian) conjugate momentum operators. The coefficients, both pre and post, in this quadratic are pure functions of the coordinate operators and are candidates for $\Theta$.

If the first two operator identities are satisfied, in an hierarchy, the so called Theta Equation (TE) is satisfied. The TE is composed only of operators that are pure functions of the coordinate operators; and it reduces to a fourth order PDE, in the position representation, with the coordinates as independent variables. It is taken to be the operator field equation of the system. The TE does not contain reference to either the linear or constant terms of the original Hamiltonian operator. These operators can take any values and are taken to be the electromagnetic potentials separate from the gravitational operators. The TE is a QM equation and includes reference to each $\Theta$ operator.

It is hypothesised (but not proved) that we do not need to go above the second operator identity, in an hierarchy, in order to discuss CM.

The relation between QM and CM, for a system, is taken to be that, in CM, *all* the operators, in the corresponding equations of QM, *commute*.

If a $\Theta$ operator, in the position representation, is defined, in turn, as the component of the fundamental tensor $g^{lm}$ of the Riemannian $n_c$ -space, then the TE becomes the Gravitational Theta Equation (GTE) defined on a flat Riemannian $n_c$ -space *C*. The dimension $n_c = n_d n_p$ where $n_d$ is the dimension of the original QM space and $n_p$ the number of particles in it. The GTE is, by the way it is formulated, *classical*.

In a Minkowski space of dimension four, Feynman/ Dyson deduced Maxwell's electromagnet equations from QM. Although this paper is not about electromagnetism (EM) be assured that Constraints Theory does not clash with EM [19]. Indeed the shape of galaxies may be due, exclusively, to EM forces; thus Dark Matter is a figment [21]. The postulates of [21] are: That stars are charged and, at the centres of galaxies, there is a magnetic dipole, in the plane of the galaxy, presumably associated with the black hole.

If the postulates of [21] are correct the shape of galaxies is due partly to gravity and partly to EM. Both, with K equation, lead to the conclusion that Dark Matter is a figment.

There is a classical curved Riemannian $n_c$ -space $C'$, associated with the space *C*, which is tangential to the space $C'$ at the points P in $C$ and $P'$ in $C'$. The space $C'$ is a general Riemannian space; and we take the curvature of $C'$ as being a symptom of gravitation. It can be proved that a Riemannian $n_c$ -space cannot be curved unless $n_c \geq 4$ .

There is a GTE which is defined on a flat Riemannian space $C$ and, according to conventional QM, aught to be expressed in Cartesian coordinates. Because it is tangential to a curved space $C'$ at the pole $\mathrm{P}'$ of Cartesian geodesics, it approximates C' over small distances.

If we take the *classical* version of the gravitational Hamiltonian (the one associated with $C'$), and we eliminate the components of momentum from Hamilton's equations, we get equations which are identical to the Geodesic Equations in a Riemannian $n_c$ -space. Both this space and everything deduced from it is classical. This, essentially, is why QM is incompatible with General Relativity.

The Kilmister Equation (the K equation) $K^a_b = 0$ is a classical tensor equation which is defined on the same space as $C'$. If we use geodesic Cartesians, with pole $\mathrm{P}'$, the K equation approximates the GTE, close to $\mathrm{P}'$, and allows $\mathrm{P}'$ to be anywhere in $C'$; the K equation is a new classical law of gravity for a particle.

The K equation is of fourth order; but all the customary second order solutions satisfy it for a gravitating particle. There are, however, extra terms. It is taken as the classical field equation when there are no other forces other than gravity.

In particular the SS (spherically symmetric) solution, to the Newtonian approximation to the K equation, has two extra SS terms in addition to Newton's inverse square law; see (7.5). These terms, if non-zero, can only be appreciable at cosmological (galactic and super galactic) distances from the source.

If, in the Newtonian scheme, we define the law of gravity

$$\nabla^2 U \approx \Lambda'; \quad |U| << 1$$

where $U$ is the dimensionless potential and compare it, with the SS solution for the approximate Newtonian K equation, for large $r$, we find that $\Lambda'$ equals four times the coefficient for $r^2$ in the solution (7.5). Since the Universe seems to be accelerating in the far field, the acceleration being proportional to distance, the coefficient $k_2$, that appears in (7.5), and $\Lambda'$ are both negative and constant.

The extra terms, in the SS Newtonian approximation to the K equation, probably account for most of the 'Dark Matter' in the halos of the galaxies (if the galaxies have halos). Due to modelling complications, however, the argument is only *suggestive*; it has not been proved. The archetypal velocity/ radius curve is satisfied by the equations of the model at the start of the plateaux; but the archetype only approximates real observations rarely.

According to [10] $G$ (Newton's constant) drifts with time. Although the drift is small the variation in $G$, due to the extra terms, is much smaller. Therefore the variation reported in [10] has nothing to do with extra terms.

If the Pioneer Anomaly is Newtonian gravitational then, according to the K equation, it is *not* simply related to the constant gravitational solar term. The balance of opinion seems to be converging on the 'small physical effect' as opposed to 'some theoretical mistake' as the culprit.

Now turning to GR: The Cosmological Metric requires the $4\times4$ $R_b^a$ (Ricci), $G_b^a$ (*Einstein*), $K_b^a$ (Kilmister) tensors to be diagonal. This means, among other things, that the expansion factor $A(u)$ of the model universe satisfies two DEs simultaneously.

In consequence the truly empty model universe (without 'Dark Energy') satisfies the Einstein's equations exactly.

In order to bring in small pressure and density we try perturbation of the empty model universe $A(u)$. This is *unsatisfactory* as it brings us back to a model universe empty of ordinary matter; (it has Dark Energy).

As a consequence of Einstein's equations and definitions we introduce Hubble's constant and Friedman's equation.

Another way of coping with the fact, that the expansion factor $A(u)$ of the model universe satisfies two DEs simultaneously, is to expand $A(u)$, about zero, with respect to the time/ distance $u$. Time/ distance equals zero is the origin 'here and now'; and our conclusions are irrespective of the sign of $u$. We actually observe $u$ negative. The expansion we have used is

$$A(u) \cong a_0 + a_1 u + a_2 u^2 + a_3 u^3 + a_4 u^4 + a_5 u^5 + O(u^6); \quad u \geq 0$$

$$A(0) \equiv 1 \Rightarrow a_0 = 1; \quad a_1 = \frac{H_0}{c}; \quad \text{Initial Conditions;} \quad 1 >> |\text{u}|$$

where $H_0$ is Hubble's constant.

In practice $a_5 = 0$; so we have at least $a_2, a_3, a_4, k, \Lambda$ to determine. The Kilmister equation $K_b^a = 0$ gives two equations that involve $k$; $G_k^j = 0$ gives two more involving $\Lambda$. The tensor equation $G_k^j = 0$, however, involves the pressure and the density. In the past $(u < 0)$ and in the future $(u > 0)$ so both pressure and density can vary from the values 'here and now'. We make the assumption that they are both linear (see (13.2)) in $u$. This is an approximation which may not, necessarily, be valid for big $u$.

The only valid method of getting more equations is to go up to level three on the hierarchies and use tensor equations; that involves a huge amount of work.

Of the nine constants $a_2, a_3, a_4, p_0, p_1, \rho_0, \rho_1, k, \Lambda$ seven are free and two presumably universal. Setting $u = 0$, for which the calculations have been done, means that $p_1 = 0, \rho_1 = 0.$ But we have at most five equations, usually four, so assumptions have to be made. For example the last calculation, for which we have five equations (the age of the model universe is $13.7\times10^{9}$ years), assumes that $k=10^{-54}$ $m^{-2}$ and $p_0=0$ gives $\Lambda=-2.08\times10^{-52}$ $m^{-2}$ and $\rho_0=4.88\times10^{-26}$ $kgm^{-3}$.

Our calculations go along way to explaining the far field expansion of the Universe on the basis of the K equation.

A.M. Deakin 7/1/2018
L. H. Kauffman 15/1/2018